\RequirePackage{lineno} 
\documentclass[%
 showpacs,
 reprint,
 amssymb,amsmath,%
 prb,aps,
floatfix
]{revtex4-1}
\usepackage[latin1]{inputenc}
\usepackage{slashbox,pict2e}

\newcommand{\eq}[1]{\begin{equation}{#1}\end{equation}}

\newcommand{\muu}{\hbox{\textmu}}
\newcommand{\ddx}{\frac{\partial}{\partial\text{x}}}
\newcommand{\ddy}{\frac{\partial}{\partial\text{y}}}
\newcommand{\ddz}{\frac{\partial}{\partial\text{z}}}
\usepackage{lmodern}
\usepackage{textcomp}
\usepackage[T1]{fontenc}
\usepackage{amsfonts}% bold math
\usepackage{amsmath}% bold math

\usepackage{graphicx}% Include figure files
\numberwithin{equation}{section}
\usepackage{cleveref}% bold math
\crefname{equation}{Eq.}{Eqs.}
\crefname{figure}{Fig.}{Figs.}
\Crefname{figure}{Figure}{Figures}
\crefname{section}{Sec.}{Secs.}
\Crefname{table}{Table}{Tables}
\crefname{table}{Tab.}{Tabs.}
\Crefname{appendix}{Appendix}{Appendices}
\crefname{appendix}{App.}{Apps.}

\usepackage{dcolumn}% Align table columns on decimal point
\usepackage{bm}% bold math
\usepackage{color}% 
\usepackage{rotating}
\usepackage{xcolor}
\usepackage{ulem}
\usepackage{xspace}%
\newcommand{\ca}[0]{Ref.~\citenum{Altmann}\xspace}
\newcommand{\gbt}[1]{\color{cyan}{}\color{black}}
\newcommand{\mad}[1]{\color{cyan}{}\color{black}}

\newcommand{\notPart}[1]{}

\newcommand{\algaas}[0]{Al$_{0.3}$Ga$_{0.7}$As\xspace}
\newcommand{\bra}[1]{\left<{#1}\right|}% <| som tilpasser seg hÃ¯Â¿Â½yden
\newcommand{\ket}[1]{\left|{#1}\right>}% |> som tilpasser seg hÃ¯Â¿Â½yden
\newcommand{\braket}[2]{\left.\bra{#1}#2\right>}
\newcommand{\norm}[2]{\left|\braket{#1}{#2}\right|^2}

\newcommand{\csv}[0]{C$_\textrm{6v}$\xspace}
\newcommand{\ctv}[0]{C$_\textrm{3v}$\xspace}
\newcommand{\dth}[0]{D$_\textrm{3h}$\xspace}
\newcommand{\dsh}[0]{D$_\textrm{6h}$\xspace}
\newcommand{\kp}[0]{\bm{$k\cdot{p}$}\xspace}
\newcommand{\tb}[0]{\\[1pt]\\}
\newcommand{\tec}[0]{PTCO\xspace}
\begin{document}
% 
% \setpagewiselinenumbers
%  \linenumbers
%\reprint{APS/123-QED}

\title{Symmetries and optical transitions of hexagonal quantum dots in GaAs/AlGaAs nanowires}% Force line breaks with \\

\author{Guro K. Svendsen}
\altaffiliation[Also at ]{University Graduate Center, Kjeller, Norway.}%Lines break automatically or can be forced with \\
\affiliation{Department of Electronics and Telecommunications, Norwegian University of Science and Technology, Trondheim, Norway}
\author{Johannes Skaar}%
\altaffiliation[Also at ]{University Graduate Center, Kjeller, Norway.}%Lines break automatically or can be forced with \\
\affiliation{Department of Electronics and Telecommunications, Norwegian University of Science and Technology, Trondheim, Norway}

\author{Helge Weman}%
\affiliation{Department of Electronics and Telecommunications, Norwegian University of Science and Technology, Trondheim, Norway}

\author{Marc-André Dupertuis}%
\affiliation{Laboratory of Physics of Nanostructures, Ecole Polytechnique Fédérale de Lausanne (EPFL), CH-1015 Lausanne, Switzerland}

%
%Authors' institution and/or address
%%\\This line break forced with \textbackslash\textbackslash
%}%

%\affiliation{
%Second institution and/or address}

\date{\today}% It is always \today, today,
       % but any date may be explicitly specified
\begin{abstract}
We investigate the properties of electronic states and optical transitions in hexagonal GaAs quantum dots within \algaas nanowires. Such dots are particularly interesting due to their high degree of symmetry. A streamlined \textit{postsymmetrization technique based on class operators} (\tec) is developed which enables one to benefit in one run from the insight brought by the \textit{Maximal symmetrization and reduction of fields} (MSRF) approach reported by Dalessi \textit{et al.}\cite{dupertuis_msr}, \textit{after} having solved the Schrödinger equation. Definite advantages of the \tec are that it does not require having to modify any existing code for the calculation of the electronic structure, and that it allows to numerically test for elevated symmetries. We show in the frame of a 4-band \kp model that despite the fact that the \dsh symmetry of the nanostructure is broken at the microscopic level by the underlying Zinc Blende crystal structure, the effect is quite small. Most of the particularities of the electronic states  and their optical emission can be understood by symmetry elevation to \dsh and the presence of approximate azimuthal and radial quantum numbers.
\end{abstract}
\maketitle
%double cite:\cite{tomioka:led,*henneghien}
\section{introduction}

Semiconductor nanowires have emerged as promising building blocks for realization of various nanoscale optoelectronic devices \cite{Lauhon20041247}. The nanowire technologies enable heterostructures with a high level of flexibility in terms of geometry and material composition \cite{Lauhon20041247}. In particular, growth of site controlled quantum dots (QDs) within nanowires show significant advantages compared to more conventional self-assembled QDs (Stranski-Krastanov)\cite{tribu,borgstrom}, and bright single photon emitters have  been demonstrated in a range of materials using nanowire QDs\cite{kats,tribu, borgstrom,choi,dorenbos}.

QDs may be used to facilitate quantum information and cryptography technologies, e.g. by emission of entangled photon pairs. In this respect nanowire QDs are especially suitable as they are highly symmetric, with a hexagonal cross section. High symmetry is a requisite for entanglement, needed in order to limit the fine structure splitting of the excitonic states which is a consequence of symmetry breaking. QDs within nanowires grown in the high symmetry direction [111] are therefore suggested as ideal sources for emission of entangled photon pairs \cite{Singh}. Cascaded emission spectra, possibly enabling such pairs have also been measured experimentally \cite{dorenbos}. Generation of entangled photons is also possible in self-assembled QDs, especially InGaAs/GaAs QDs grown on [111] substrates are proposed ideal for such generation, due to the threefold rotation symmetry of this surface\cite{Schliwa}. QDs may also enable quantum computation by using the electron spin as a quantum bit \cite{imamoglu}, QDs grown in inverted pyramids on [111] GaAs substrates \cite{Oberli} are also promising in this respect.

With increasing control of heterostructure shape and material composition, comes the possibility to grow structures according to optimized design. Theoretical models e.g. providing insight into the nature of electronic states are essential in such optimization. In this work we calculate the electronic states and optical transitions in hexagonal GaAs/AlGaAs nanowire QDs, and analyze the results taking advantage of their high symmetry. A detailed procedure for full-depth symmetry analysis and reduction of computational domain has been presented previously by Dalessi \textit{et al.}\cite{dupertuis_msr}. Here we will pursue this effort by developing a new procedure alleviating recoding: a postsymmetrization technique using class operators (\tec). The \tec is very general and applies independently of the method employed for the calculation of the electronic structure (\kp, tight-binding e.t.c). It also provides a systematic and flexible procedure to test possible elevated symmetries.

In \cref{sec:kp_calc} relevant theoretical studies of similar semiconductor heterostructures are considered, and we present the general features of the \kp model used in this paper. 
The explicit QD under consideration is described in \cref{sec:symmetry}, and the symmetries of our model structure are identified, providing the premises to optimally choose the basis of our \kp Hamiltonian. Having fully specified the numerical model, we describe the pertaining symmetry implications on the QD eigenstates in \cref{sec:nwqd_env_fct}. The \tec is then presented in \cref{sec:postsymmetrization}.  \Cref{sec:calc_c3v,sec:opt_trans} contain the numerical calculations,  analyzed using the \tec. We show in \cref{sec:calc_c3v} that the analysis gives rise to a deeper understanding of the electronic states, in particular of the level sequences. In \cref{sec:opt_trans} we investigate the fine structure of the spectrum of squared momentum matrix elements. We prove that symmetry elevation to \dsh and the existence of azimuthal and radial quantum numbers are necessary ingredients to explain many missing/weak transitions.

%\section{Quantum dot description and \texorpdfstring{\kp}{kp} model}
\section{Quantum dot description and \kp model}
\label{sec:kp_calc} 
Despite large activity within the experimental realization of nanowire QDs, there has until now been much less attention towards numerical calculations of the electronic states. \textit{Niquet et al.}\cite{Niquet} did perform calculations of strained InAs/InP  nanowire QDs, using a tight binding model. The optical transitions were given, and labeled using group symmetry. The QDs were however approximated as cylindrical in that work;  cylindrically shaped QDs grown in the [111] direction of a wurtzite structure will inherit the \ctv symmetry of the crystal. Of related interest is also the work of \textit{Zhang et al.}\cite{zhang_2011}, considering excitons in nanowire QDs in the strained InGaN/GaN material system, using an effective mass approximation. Nanowires in the GaAs/AlGaAs material system considered in the current paper were treated by \textit{Kishore et al.} \cite{Kishore} using the \kp model, but no calculations exists, to our knowledge, of nanowire QDs within this material system. 

The \kp theory, originally intended for the calculation of band structure of crystalline solids, has also been widely used for calculation of band structures in heterostructures including QDs. Large emphasis has been on the strained self assembled QDs \cite{andreev,vukmirovic,tomic,pryor,mlinar_apl_2007}, also including calculations on GaAs/AlGaAs QDs \cite{mlinar_apl_2007}. 
   
QDs with the particular hexagonal shape considered here were studied numerically using the \kp model in  Ref.~\citenum{andreev}, no explicit usage was however made therein of the symmetry properties. Ref.~\citenum{tronc} present a strictly qualitative symmetry study of electronic states and optical transitions in hexagonal QDs.  

In the \kp model one assumes a weak interaction, \kp, between the crystal momentum and the electron momentum  \cite{LuttingerKohn,Fishman}. Using Bloch's theorem, the wavefunction can be separated into a slowly varying envelope function,  and a rapidly varying part with the periodicity of the crystal lattice.  Note that this approximation is only valid close to the zone center ($\mathbf{k}=0$), and it is not capable of describing an interface between materials of different crystal structures, e.g. a transition from a Wurtzite to a Zinc Blende material. 

The symmetry preserving ability of the \kp model has been investigated numerically\cite{tomic_vukmirovic}, demonstrating that the true symmetry of any structure can be restored upon inclusion of enough bands and interface terms in the \kp model. Care should nonetheless be taken to distinguish between the symmetry of any simplified numerical model and the physical system.

We shall use a simple \kp model to describe the states of a GaAs QD within an AlGaAs nanowire. The electrons of the conduction band is described using an effective mass model, and the holes of the valence band are described using a 4-band Luttinger Hamiltonian. This rather simple model will be used to demonstrate the PTCO, which can also be used in more complex theoretical frameworks.
  
The conduction band describes electrons with spin $j=1/2$, the electrons can however be written using a scalar Schrödinger equation with an effective mass approximation, ignoring mixing with other bands \cite{dupertuis_msr}
\begin{equation}
H=-\frac{\hbar^2}{2m_0}\nabla\frac{1}{m^*(\mathbf{r})}\nabla+V_{\textrm{cb}}(\mathbf{r}).
\label{eq:cb_ham}
\end{equation}
Here, $\nabla$ is the 3D differential operator $\nabla=\ddx \mathbf{u_x}+\ddy \mathbf{u_y}+\ddz \mathbf{u_z}$, $m^*(\mathbf{r})$ is the effective electron mass in units of the electron mass $m_0$, and $V_{\text{cb}}(\mathbf{r})$ is the effective confinement potential for electrons in the conduction band. The envelope function $\psi_n$ of energy level $n$ is given by the Schrödinger equation:

\begin{equation}
H \psi_n=E_n\, \psi_n.
\label{eq:cb_schr}
\end{equation}
The spinorial nature of the conduction band states can be restored later; the exact procedure is given in \cite{dupertuis_opt}.

Band mixing and spin cannot be ignored for the holes. The top six valence bands can be described as multiplet states with spin $j=3/2$, and  $j=1/2$.\cite{LuttingerKohn} The latter multiplet is the 
split-off band, separated from the first multiplet with the amount $\Delta_{\textrm{so}}$. If coupling to this split-off band can be ignored, we are lead to a four 
valence band \kp envelope function model. We can use the 4x4 envelope function Luttinger Hamiltonian for diamond, if we neglect in addition inversion symmetry breaking in GaAs/AlGaAs. This can always be expressed in the 
form \cite{Fishman,LuttingerKohn}.
\begin{equation}
H=\frac{-\hbar^2}{m_0}\left(\begin{array}{cccc}
p+q&-s&r&0\\
-s^+&p-q&0&r\\
r^+&0&p-q&s\\
0&r^+&s^+&p+q\end{array}\right)+V_{\text{vb}}(\mathbf{r}).
\label{eq:vb_ham}
\end{equation}
Here, $p,q,r,s$ are second order polynomials of differential operators. Their exact expressions depend on the Bloch basis which will be chosen later after considering the symmetry of the Hamiltonian. The polynomial coefficients are given in terms of the Luttinger parameters $\gamma_i(\mathbf{r}), i=1\dots3$, and $V_{\text{vb}}(\mathbf{r})$ is the confinement potential. The valence band spinors $\uline\psi_n$ are found using \cref{eq:cb_schr,eq:vb_ham}. The underline will be used throughout to distinguish spinors from scalar functions.

\section{Model structure and its symmetry}
\label{sec:symmetry}
In this section we identify the symmetries of our model structure, introduce some concepts from group theory, and review its implications for the best Bloch basis. This basis will be given from the Maximal 
symmetrization and reduction of fields (MSRF) technique, previously presented by Dalessi\textit{ et al.} \cite{dupertuis_msr}.

\begin{figure}[htbp]
	\includegraphics[width=\columnwidth]{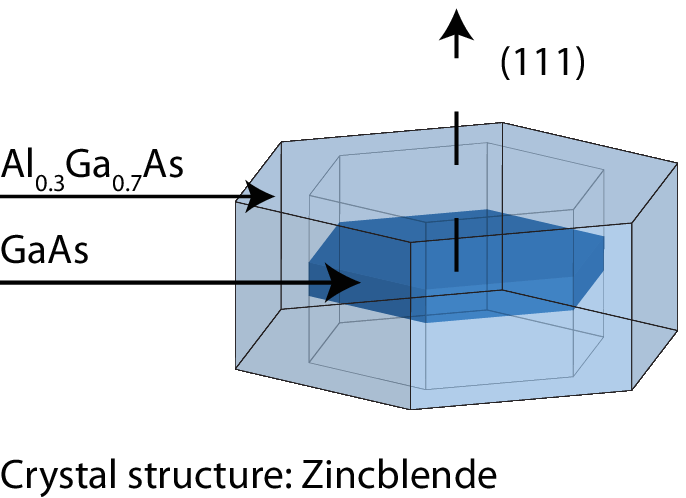}
	\caption{Schematics of a GaAs quantum dot within an \algaas nanowire}
	\label{fig:structure}
\end{figure}

The model structure considered in this paper is a QD grown as an axial insert of GaAs within an \algaas nanowire. A radial shell of \algaas is grown around the dot so that it is surrounded by \algaas in all 
directions. The growth direction defining the nanowire axis is the crystal direction [111], and the cross-section is hexagonal. We will assume a Zinc Blende crystal structure. A schematic of the 
structure is shown in Fig.~\ref{fig:structure}. Similar QD structures have been grown previously by \textit{Kats et al.}\cite{kats}, however, they did obtain mixed crystal phases containing both wurtzite and Zinc Blende. Pure Zinc Blende GaAs/AlGaAs axial heterostructures has however been realized \cite{Guo2011}. 

The true QD symmetry is the common symmetry of the mesoscopic heterostrocture and the microscopic crystal structure. The mesoscopic structure (\cref{fig:structure}) is symmetric (i.e. invariant) under discrete $\pm2\pi/6$ rotations and under mirror operations w.r.t the six vertical planes containing the rotation axis, as well as w.r.t. the horizontal plane orthogonal to the rotation axis. These symmetry operations and their compositions form a point group of 24 elements called \dsh\cite{Altmann}. Mathematically, the mesoscopic symmetry restriction on the conduction band Hamiltonian is expressed by the invariance relation pertaining to the confinement potential and effective mass appearing in \cref{eq:cb_ham}:

\begin{equation}
\left.\begin{array}{c}
V_\text{cb}(\mathbf{r})=V_\text{cb}(\mathcal{R}(g)^{-1}\mathbf{r})\\
m^*(\mathbf{r})=m^*(\mathcal{R}(g)^{-1}\mathbf{r})\end{array}\right\} \forall  g\in \textrm{\dsh}.
\label{eq:inv_rel_cb}
\end{equation}	  
Here, $\mathcal{R}$ is a set of standard representation matrices operating in 3D space \cite{Messiah,dupertuis_msr}, and $\mathcal{R}(g)$ represents change of coordinates indexed by the group element $g$. The symmetry of our conduction band Hamiltonian (\cref{eq:cb_ham}) is then directly given by the mesoscopic symmetry \dsh. %on $V_\textrm{cb}(\mathbf{r})$is also given by 

\begin{figure}[htbp]
	\includegraphics[width=\columnwidth]{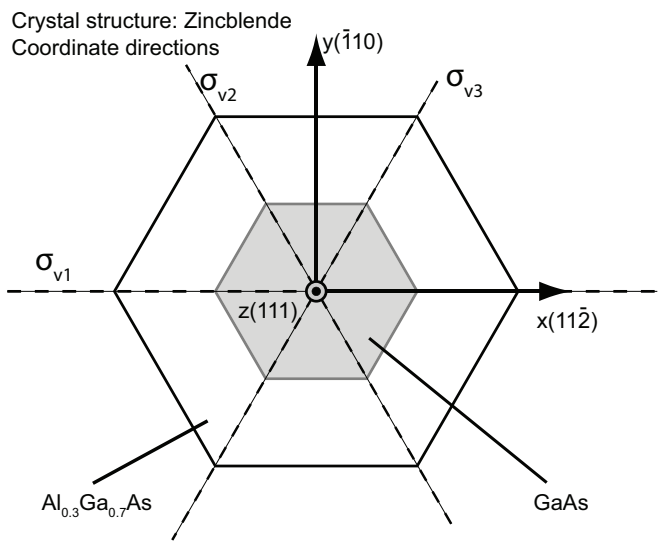}
	\caption{Coordinate axes and crystal directions for the hexagonal QD.}
	\label{fig:koord_dir}
\end{figure}

For the valence band, there are similar invariance constrictions due to the mesoscopic symmetry on  $V_\textrm{vb}(\mathbf{r})$, and also on the spatially dependent Luttinger parameters $\gamma_i(\mathbf{r})$ (appearing in the $p,q,r,s$ polynomials) 
\begin{equation}
\left.\begin{array}{c}
V_\textrm{vb}(\mathbf{r})=V_\textrm{vb}(\mathcal{R}(g)^{-1}\mathbf{r})\\
\gamma_i(\mathbf{r})=\gamma_i(\mathcal{R}(g)^{-1}\mathbf{r}), i=1,2,3\end{array}\right\} \forall  g\in \textrm{\dsh}.
\label{eq:inv_rel_vb}
\end{equation}	  
However, the valence band Luttinger Hamiltonian (\cref{eq:vb_ham}) also carries a face centered cubic diamond $O_h$ symmetry due to the underlying crystal, contained within the first term of \cref{eq:vb_ham}. When the orientation of the crystal axes w.r.t the mesostructure is as in \cref{fig:koord_dir}, the common symmetry elements are those of the group \ctv, i.e. $\{e,C_3^+,C_3^-,\sigma_{v1},\sigma_{v2},\sigma_{v3}\}$. Here $C_3^+$ and $C_3^-$ are discrete $2\pi/3$ rotations and $\sigma_{vi},i=1\ldots3$ are three vertical mirror operations shown in Fig.~\ref{fig:koord_dir}. Hence the valence band Hamiltonian (\cref{eq:vb_ham}) has only \ctv symmetry. 
In the following we will therefore use the \ctv group as the main reference when introducing the necessary group theory concepts.

Description of the group, including group multiplication tables and group representations can be found in reference books like Altman\cite{Altmann}. For the single group \ctv there are three irreducible representations (irreps), $E$, $A_1$ and $A_2$. The $E$ irrep is a 2D representation whilst the $A_i$ irreps are 1D.

So far we have just considered the single group, describing operations performed on the spatial coordinates. The spinorial nature of the conduction and valence band does however necessitate the use of 
a double group representation\footnote{Note that in spin space a rotation by $2\pi$ is equal to inversion, and a rotation of $4\pi$ define the identity operation, $E$. The double group therefore has twice the number of elements compared to the single group.}. In the \ctv double group there are two 1D irreps $^1E_{3/2}$ and $^2E_{3/2}$ and one 2D irrep $E_{1/2}$. \cite{Altmann}. With spin a change of the spatial coordinates $\vartheta(g)^{(3D)}$ ($\vartheta(g)^{(3D)}$ is an abstract functional operation corresponding to the change of coordinates described by the matrix $\mathcal{R}(g)$) should be accompanied by a corresponding change in spin space, $\vartheta(g)^{(j)}$. Such composite operations are thus defined by $\vartheta(g)=\vartheta(g)^{(3D)}\otimes\vartheta(g)^{(j)}$, where $\otimes$ is the tensor product between both operator spaces, and $g$ becomes an element of the double group. A natural choice for  a standard representation of rotations in spinorial space in 
terms of Euler angles, $g\equiv(\alpha,\beta,\gamma)$, is the set of 4x4 Wigner matrices, $W(\alpha,\beta,\gamma)$\cite{Messiah}, since the Luttinger Hamiltonian is usually expressed in a Bloch function basis 
transforming like angular momentum and indexed $\ket{j,m}$, $j=3/2$, where $m$ is the component along some quantization axis. Any matrix for improper rotation (i.e. mirror operations) can be obtained 
as a combination of proper rotations and spatial inversion, $i$, but since the diamond crystal structure is even under $i$, this can be ignored.

To fully take advantage of the symmetry properties of the valence band Hamiltonian, a specific choice of the Bloch basis is needed \cite{dupertuis_msr}. For some low symmetry groups, an optimal quantization axis can be found \cite{dupertuis_msr}, enabling full depth symmetry analysis. However, as explained in Ref.~\citenum{dupertuis_msr}, in general a rotation of the quantization axis is not sufficient, and a more general unitary transformation of the basis has to be performed. The elements of the optimum basis are called \cite{dupertuis_msr} \textit{Heterostructure Symmetrized Bloch Functions} (HSBF's), and are labeled by irreps of the double group. 

The HSBF's are symmetrized superpositions of the usual Bloch functions. For our \ctv Hamiltonian, the expressions for the HSBF's are given by \cref{eq:vb_hsbf_dsh_opt,eq:vb_hsbf_dsh_msr} of \cref{app:dsh_hsbf}. The HSBF basis is written in terms of them as 
\begin{equation}
\{\ket{^1E_{3/2}},\ket{E_{1/2},1},\ket{E_{1/2},2},\ket{^2E_{3/2}}\}.
\label{eq:HSBF}
\end{equation}
The ordering in \cref{eq:HSBF} is important to preserve the form of the time reversal symmetry operator \cite{dupertuis_opt}. 

Note that our choice of axes (\cref{fig:koord_dir}) differs from that of Refs.~\citenum{dupertuis_msr,dupertuis_opt}, in which the $z$-axis in the equations for the HSBF's pertain to a quantization axis in the crystal direction $[\bar{1}10]$, i.e. corresponding to the $y$-axis in our present choice of coordinates. We have rather chosen the $z$-axis based on the special axis of the nanostructure ([111]), as the appropriate concept of light hole (LH) and heavy hole (HH) for our structure relates to $m_z=\pm\frac{1}{2}$ or $m_z=\pm\frac{3}{2}$ respectively for a $z$-axis along [111].%as the concept of quantization axis is not anymore important using the HSBF basis. 

In the HSBF basis, the $p,q,r,s$ polynomials appearing in \cref{eq:vb_ham} are given by

\begin{subequations}
\label{eq:pqrs}
\begin{align}
&p=-\frac{1}{2}\left(\frac{\partial}{\partial{x}}\gamma_1\frac{\partial}{\partial{x}}+\frac{\partial}{\partial{y}}\gamma_1\frac{\partial}{\partial{y}}+\frac{\partial}{\partial{z}}\gamma_1\frac{\partial}{\partial{z}}\right)\label{eq:p}\\
&q=\frac{1}{2}\left(-\frac{\partial}{\partial{x}}\gamma_3\frac{\partial}{\partial{x}}-\frac{\partial}{\partial{y}}\gamma_3\frac{\partial}{\partial{y}}+2\frac{\partial}{\partial{z}}\gamma_3\frac{\partial}{\partial{z}}\right)\label{eq:q}\\
&r=-\frac{\partial}{\partial{x}}b\frac{\partial}{\partial{x}}+\frac{\partial}{\partial{y}}b\frac{\partial}{\partial{y}}-\frac{\partial}{\partial{z}}a\frac{\partial}{\partial{x}}-\frac{\partial}{\partial{x}}a\frac{\partial}{\partial{z}}\\
&s=-\frac{\partial}{\partial{y}}a\frac{\partial}{\partial{z}}-\frac{\partial}{\partial{z}}a\frac{\partial}{\partial{y}}+\frac{\partial}{\partial{x}}b\frac{\partial}{\partial{y}}+\frac{\partial}{\partial{y}}b\frac{\partial}{\partial{x}}
\end{align}
\end{subequations}

\begin{subequations}
\label{eq:ab}
\begin{align}
a=\frac{\epsilon}{2}\left(-\gamma_3-i\sqrt{2}\gamma_2\right)\\
b=\frac{i\epsilon}{2}\left(\gamma_2+i\sqrt{2}\gamma_3\right)
\end{align}
\end{subequations}
with $\epsilon=\frac{1-i\sqrt{2}}{\sqrt{3}}$. These expressions have been obtained by using the bulk expressions, e.g. by taking the Luttinger Hamiltonian expressed in direction [111] in Ref.~\citenum{Fishman}, then changing its basis to the HSBF given in \cref{{app:dsh_hsbf}}, and finally by replacing $k_j\to -i\frac{\partial}{\partial{j}}$ for $j=x,y,z$.

In the forthcoming analysis, it will be useful to describe the spinorial nature of the quantum states using the concept of light hole (LH) and heavy hole (HH) dominant character (because of bandmixing the HH and LH states are mixed). 
Raw approximations for the effective masses parallel and perpendicular to the [111] axis can be read from \cref{eq:p,eq:q} (assuming $r$=$s$=0), and are given by
\begin{subequations}
\label{eq:eff_mass}
\begin{align}
m^*_{||}=\frac{1}{\gamma_1\mp2\gamma_3}\\
m^*_{\perp}=\frac{1}{\gamma_1\pm\gamma_3},
\end{align}
\end{subequations}
where the upper and lower sign applies to HH and LH respectively. For a normalized state $\ket\psi$, the weight of the LH contribution is defined by

\begin{equation}
w^{[111]}_{LH}(\psi)\equiv\left|_z\langle3/2,1/2|\psi\rangle\right|^2+|_z\langle3/2,-1/2|\psi\rangle|^2,
\label{eq:lh_weight}
\end{equation}
and similarly for the HH, so $w^{[111]}_{LH}+w^{[111]}_{HH}=1$. Using the basis change between the HSBF basis and the basis with quantization axis $[111]$, it is easy to show that
\begin{equation}
w^{[111]}_{LH}(\psi)=\left|\braket{E_{1/2},1}{\psi}\right|^2+\left|\braket{E_{1/2},2}{\psi}\right|^2.
\label{eq:lh_weight_hsbf}
\end{equation}
We thus note that LH/HH character w.r.t. [111] is conveniently in one to one correspondence with $E_{1/2}/^iE_{3/2}$ character in the \ctv HSBF.

\section{Symmetry implications on nanowire quantum dot states}
\label{sec:nwqd_env_fct}
The HSBF basis allows to maximally symmetrize the envelope functions appearing in the valence band spinors, and to give interpretation of the spinors as dominant product states. These possibilities, which are consequences of symmetry, are explored in the forthcoming subsections.
\subsection{Symmetry of envelope functions}
%\label{sec:nwqd_env_fct}

We first recall the transformation properties of the double group spinors, in order to study the properties of envelope functions occurring in our nanowire QD. This is based on the MSRF technique\cite{dupertuis_msr}. The HSBF-transformed Wigner matrices are denoted by $V^B(g)$, as they are block matrices in this basis\cite{dupertuis_msr}. Indeed, one can write them as a direct sum, $^1E_{3/2}\oplus E_{1/2}\oplus^2E_{3/2}$, of the irrep matrices of the \ctv double group \cite{dupertuis_msr}. Therefore the spinor transformation (passive point of view), under any double group operation $g$,  can then be written as

\begin{equation}
\left[\uline{\psi}(\mathbf{r})\right]'_j=\left[\vartheta(g)\uline{\psi}(\mathbf{r})\right]_j=\sum_k [V^B(g)]_{j,k}[\uline{\psi}(\mathbf{\mathcal{R}}(g)^{-1}\bm{r})]_k.
\label{eq:general_transf}
\end{equation}
A seminal consequence of symmetry in a system subjected to invariance relations like \cref{eq:inv_rel_cb,eq:inv_rel_vb} is that it is always possible \cite{Hammermesh} to label an eigenstate $\uline{\psi}$ by an irrep, $\Gamma$, and a partner function index, $\mu$ of the symmetry group, such that its corresponding transformation law is also:

\begin{equation}
\vartheta(g){\underline{\psi}}^{{\Gamma}}_{\mu}(\mathbf{r})=\sum_{\nu=1}^{d_{{\Gamma}}}\left[D^{{\Gamma}}(g)\right]_{\mu,\nu}\underline{\psi}^{\Gamma}_\nu(\mathbf{r})
\label{eq:group_ip_irrep}
\end{equation}
where $D^{{\Gamma}}(g)$ is a set of representation matrices for $\Gamma$, and $d^\Gamma$ is the dimension of $\Gamma$. For $D^{{\Gamma}}(g)$ we shall use the matrices of Refs.~\citenum{dupertuis_msr,Messiah}, which correspond to a transposed multiplication table w.r.t. Ref.~\citenum{Altmann} since we use the passive point of view.

It is clear that \cref{eq:general_transf,eq:group_ip_irrep} strongly constrain the envelope function shapes, and it was shown in \cite{dupertuis_msr} that they can be uniquely decomposed into \textit{ultimately reduced envelope function} (UREF) components, transforming according to single group irreps. The classification of UREFs using single group labels could be used to reduce the computational domain significantly as argued in Ref.~\citenum{dupertuis_msr}. We will see in the present work that it is also very helpful in the postprocessing and analysis of the data. 
In our QD, the full valence band spinors can be labeled by \ctv double group irreps, and expressed in terms of UREFs as in \cref{eq:vb_spinors}

\begin{subequations}
\label{eq:vb_spinors}
\begin{align}
\underline{\psi}^{^1E_{3/2}}=\left(\begin{array}{c}
\phi^{A_1}\\
-\phi^{E}_2\\
\phi^{E}_1\\
-\phi^{A_2}\\	
\end{array}\right)\label{eq:psi_1_E_32}\end{align}\begin{align}
\underline{\psi}^{^2E_{3/2}}=\left(\begin{array}{c}
\phi^{A_2^*}\\
\phi^{E^*}_1\\
\phi^{E^*}_2\\
\phi^{A_1^*}\\	
\end{array}\right)\label{eq:psi_2_E_32}\end{align}\begin{align}
\underline{\psi}^{E_{1/2}}_1=\left(\begin{array}{c}
-\phi^{E}_2\\
\frac{1}{\sqrt{2}}[\phi^{A_1}+\Phi^E_1]\\
-\frac{1}{\sqrt{2}}[\phi^{A_2}+\Phi^E_2]\\
\varphi^{E}_1\\	
\end{array}\right)\label{eq:psi_E_12_1}\end{align}\begin{align}
\underline{\psi}^{E_{1/2}}_2=\left(\begin{array}{c}
\phi^{E}_1\\
\frac{1}{\sqrt{2}}[\phi^{A_2}-\Phi^E_2]\\
\frac{1}{\sqrt{2}}[\phi^{A_1}-\Phi^E_1]\\
\varphi^{E}_2\\	
\end{array}\right)\label{eq:psi_E_12_2}.
\end{align}
\end{subequations}
The envelope functions have here been labeled in a simplified manner, keeping the double group labels of the global spinor and of the HSBF\cite{dupertuis_opt} implicit. The full labels should be restored for description of how the envelope functions of separate subequations in \cref{eq:vb_spinors} transform into each other. 

It should be pointed out that for QDs time reversal induces a unique mapping between Kramers degenerate spinors, in analogy to the $k\to-k$ mapping of quantum wires \cite{dupertuis_opt}. The time reversal operator, $K$, is \cite{dupertuis_opt} $K=FK_0$, where $K_0$, is a complex conjugate operator and $F_{j,k}=\delta_{k,(5-j)}(-1)^j , \,\,(j,k)=1\ldots4$. The 2D $E_{1/2}$ irrep is self conjugated, and the 1D irreps are mutually conjugated. Therefore, the pair of eigenstates belonging to a given eigenvalue is either the two partners of the self conjugated 2D irrep ($E_{1/2}$) or the pair of mutually conjugated 1D irreps ($^1E_{3/2}$, $^2E_{3/2}$). Accordingly, $K\underline{\psi}^{^1E_{3/2}}=\underline{\psi}^{{}^2E_{3/2}}$ and $K\underline{\psi}^{E_{1/2}}_2=\underline{\psi}^{E_{1/2}}_1$. It is easy to see from the form of $K$ that the UREFs appearing in \eqref{eq:psi_1_E_32} must then be equal to the corresponding functions in \eqref{eq:psi_2_E_32}. Similarly, for the ${E_{1/2}}$ irrep one obtains restrictions on \eqref{eq:psi_E_12_1} and \eqref{eq:psi_E_12_2}:
\begin{align}
\left.\begin{array}{l}
\varphi^E_i={\phi^E_i}^*\\
\phi^{A_i}={\phi^{A_i}}^*\Rightarrow\textrm{$\phi^{A_i}$ is real}\\
\Phi^E_i=-{\Phi^E_i}^*,\Rightarrow\textrm{$\Phi^E_i$ is imaginary}\end{array} \right\}, \,\,\,i=1,2.
\label{eq:time_rev_req}
\end{align}

In the scalar approximation, \ctv conduction band envelope functions can be given single group labels directly and states are either non-degenerate, $\Gamma=A_i, i=1\ldots2$, or twice degenerate, $\Gamma=E$. Accounting for spin, time reversal symmetry implies Kramers degeneracy and the eigenstates of the conduction band are either two-fold or four-fold degenerate. This also holds when one considers the full \dsh symmetry of the conduction band. 

Having established in the present section the precise symmetry of every spinorial component using UREFs bearing single group irrep labels, we now propose to quantify the respective contributions of each irrep, assuming that the full spinors are all normalized to unity, i.e. $\left\|\uline{\psi}_j\right\|=1$.

The set of UREFs appearing in a given set $\uline{\psi}^\Gamma_\mu(\mathbf{r})$, transforming into each other according to  $D^\Gamma(g)$ (of \cref{eq:group_ip_irrep}) are specified by $\psi^{\Gamma,\Gamma_a}_{\Gamma_b,\mu_a}$. Here, $\Gamma_b$ label the relevant HSBF block and ($\Gamma_a, \mu_a$) every individual UREF. There are internal redundancies within the UREFs due to their own transformation properties. We can define the weight of a subset of symmetry $\Gamma_a$ within the block $\Gamma_b$ of the spinor $\uline\psi^\Gamma_\mu$ (assuming that $\Gamma_b$ occurs only once in $\Gamma$) \cite{gallinet_photonic}:

\begin{equation}
w^{\Gamma_b,\Gamma_a}\left(\uline\psi^\Gamma_\mu\right)=\left\|{\psi}^{\Gamma,\Gamma_a}_{\Gamma_b}\right\|
\label{eq:weight_uref}
\end{equation}
is independent of $\mu$. In addition $\left\|{\psi}^{\Gamma,\Gamma_a}_{\Gamma_b}\right\|=\left\|{\psi}^{\Gamma,\Gamma_a}_{\Gamma_b,\mu_a}\right\|$ is independent of $\mu_a$ by symmetry\cite{gallinet_photonic} (generalized Wigner-Eckart theorem); therefore partner function indices $\mu$ and $\mu_a$ are not relevant. With the chosen normalization, one of course has the completeness $\sum_{\Gamma_b,\Gamma_a}d_{\Gamma_a}w^{\Gamma_b,\Gamma_a}=1$. The redundancy by $d_{\Gamma_a}$ is because every UREF partner function contribute equal weight.

\subsection{Product states, dominant HSBF as DPGPS}
\label{sec:hsbf_dpgps_theory}
In a \kp model, the global symmetry of the eigenstates of the heterostructure can always be expanded into sums of products between symmetrized scalar envelope functions labeled by single group $\Gamma$,$\mu$ (UREFs), and a HSBF labeled by double group $\Gamma$,$\mu$. Often, when band mixing is not too large, there is one dominant term in the expansion. One can infer that the first factor describes well the spatial probability distribution, and the second factor entails the symmetrized spin information encoded in the HSBF. 
In such a case we will denote the relevant HSBF the \textit{Discrete Point Group Pseudo Spin} (DPGPS)\cite{Dupertuis_letter}. This notation express the fact that one can add such spinor parts in the same way as ordinary spin, but using the Wigner-Eckart theorem for point groups. Expressing the valence and conduction band states as such product states is often useful e.g. when building excitonic and biexcitonic complexes, and for simple understanding of degeneracies and other properties \cite{Dupertuis_letter}. As an example, consider an exciton with \ctv symmetry. Assume that the symmetry of the main electron states contributing to this exciton can be written as a direct product $\Gamma^\textrm{electron}=A_{1}\otimes{E_{1/2}}$. Similarly, assume that the symmetry of the hole states can be written as $\Gamma^\textrm{hole}=A_{1}\otimes\left({^1E_{3/2}\oplus^2E_{3/2}}\right)$. The excitonic states can then be constructed by combining separately the spinorial parts and the envelope parts. The symmetry of the excitonic states is thus $\Gamma^\textrm{ex}=\left(A_{1}\otimes{A_1}\right)\otimes\left({E_{1/2}}\otimes{\left({^1E_{3/2}\oplus^2E_{3/2}}\right)}\right)=2{E}$.

For the conduction band, writing the states as product states is straightforward. It is more complicated for the valence band, as band mixing may prevent a description of the states in terms of 
individual DPGPS. In the HSBF basis it is however easy to test the weight of every UREF part appearing in Eq.~\eqref{eq:vb_spinors} with the help of projection operators for scalar functions (on single group irreps) and calculating the weight (\cref{eq:weight_uref}) of every individual component. We shall see that it often leads to interpretation of the nature of the eigenstates as product states of a dominant envelope function with a corresponding DPGPS. For example an $^1E_ {3/2}$ state \eqref{eq:psi_1_E_32} with the dominant weight in the first ($A_1$) component, corresponds to the product state 
$\Gamma^\textrm{hole}=A_{1}\otimes^1E_{3/2}$.
 
One last comment regarding the DPGPS and its relationship with the HH/LH concept: as explained in \cref{sec:symmetry}, in \ctv the HH-weight equals the combined weight of the two $E^i_{3/2}$ spinor components. The valence band state of the previous example (typically our QD ground state) may therefore be classified as HH-like. As shown in \cref{app:uref_dsh} the two corresponding "spin states" are distinct conjugated DPGPS in \ctv, and to the two partners of $E_{3/2,g}$ DPGPS in \dsh. Similarly LH-like spin states will be associated to the partners of $E_{1/2}$ DPGPS in \ctv and to those of $E_{1/2,g}$ DPGPS in \dsh.

\section{Postsymmetrization}\label{sec:postsymmetrization}
Postsymmetrization has been first demonstrated by \textit{Gallinet et al.} using projection operators \cite{gallinet_photonic}. We will now present the PTCO, a novel systematic and streamlined postsymmetrization method developed 
to disentangle and classify numerical eigenstates such that they can be given individual symmetry labels. The advantages of the PTCO are: 1) that it can be used in combination with possibly existing calculation code for the electronic structure,  so that little additional numerical/theoretical work is necessary (it is general and not limited to \kp theories). 2) That it will allow to test and qualify possible approximate elevated symmetries. We stress again that though these techniques are here exemplified starting from a simple 4 band \kp model, the PTCO can equally well be applied in the frame of more complex models. 

For postsymmetrization, one considers a subspace of the space spanned by the set of eigenstates of the Hamiltonian. This subspace $\mathcal{S}$ of dimension $d_\mathcal{S}$ is obtained after selection of a few relevant eigenstates, $\{\uline{\psi}
_j, j=1\ldots d_\mathcal{S}\}$, which are numerical solutions of \cref{eq:cb_ham} or \cref{eq:vb_ham}. It will be called the \textit{solution space}. Usually we shall retain a finite set of neighboring eigenstates, from the ground state to a given maximum energy, e.g. $d_\mathcal{S}$=20 in \cref{sec:calc_c3v}. We shall assume that the states $\uline{\psi}_j$ are not entirely as symmetric as they should be (due to numerical errors or due to the use of a non-symmetric grid). Postsymmetrization techniques will aim at finding the best symmetrized states that can be found in $\mathcal{S}$, on the basis of the ideal symmetry properties that such linear combinations should satisfy. 

For any operator, $F$, there is a matrix representation in $\mathcal{S}$, called $O(F)$, defined as
\begin{equation}
O(F)_{i,j}=\braket{\uline\psi_i}{F\uline\psi_j}.
\label{eq:big_d}
\end{equation}
In the following subsections all operators, including the Hamiltonian $H$, the symmetry operators $\vartheta(g)$, the projection operators $P^\Gamma_\mu$, and the class operators $C_i$, will be identified with their matrix representation in $\mathcal{S}$. 

Note that in the general case numerical eigenstates do not exactly obey \cref{eq:group_ip_irrep} and they can be mixed in two ways. First, mixing can be due to fundamental degeneracy and the arbitrary choice of basis by the solver within every eigenspace. This effect can be strong but is not in violation of the symmetry of the (analytical) Schrödinger equation. Disentanglement is however necessary to enable the use of individual group labels corresponding to standard irreps $D^\Gamma(g)$ and belonging partner function indices $\mu$. Second, eigenstates can be mixed due to numerical inaccuracies, in particular if the grid does not respect the symmetry. Such mixing is always weak, when good convergence is achieved. If energy levels are closely spaced, symmetry breaking due to the grid may nevertheless couple states at nearby energies and with different symmetries. %To take both kinds of mixing into account, we have developed a more 
The goal of the \tec, which is a global postsymmetrization method, is to disentangle both kinds of mixing simultaneously and it will be presented after a short description of a more standard technique based on projection operators.

\subsection{Postsymmetrization using projection operators}
A standard procedure to analyze and postsymmetrize the computed basis $\{\uline{\psi}
_j, j=1...d_\mathcal{S}\}$ is to use the a set of projection operators, $P_{\mu}^{\Gamma}$ which project any state onto the partner function $\mu$ of the irrep $\Gamma$. The projection operators are given by \cite{Hammermesh}

\begin{equation}
P_{\mu}^{\Gamma}=\frac{d_{\Gamma}}{|\mathcal{G}|}\sum_{g \in \mathcal{G}}{\left({D^{\Gamma}(g)}_{\mu \mu}^*\vartheta(g)\right)}
\label{eq:projector}
\end{equation}
where $|\mathcal{G}|$ is the number of elements of the group $\mathcal{G}$. A matrix representation in $\mathcal{S}$ can be built using \cref{eq:big_d}. The set of projection operators $P_{\mu}^{\Gamma}$ are orthogonal, and obeys
\begin{equation}
\sum_{\Gamma,\mu}P_{\mu}^{\Gamma}=1.
\label{eq:projector_completeness}
\end{equation}
They can uniquely decompose any state into 
\eq{
\uline\psi=\left\{\sum_{\Gamma,\mu} P_{\mu}^{\Gamma}\right\}\uline\psi=\sum_{\Gamma,\mu} \uline\psi_{\mu}^{\Gamma}.
\label{eq:decomp_irreps}
}

The fact that the grid slightly breaks invariance will break the expected symmetry properties of the computed eigenstates $P_{\mu}^{\Gamma}\psi_{\mu'}^{\Gamma'}=\delta_{\Gamma,\Gamma'}\delta_{\mu,\mu'}\psi_{\mu}^{\Gamma}$. To quantify the amount of symmetry breaking for the basis functions of each double group irrep, we define the weight of $(\Gamma,\mu)$ in spinor $\uline\psi$ by 

\begin{equation}
w^\Gamma_\mu\left(\uline\psi\right)=\left\|P_{\mu}^{\Gamma}\uline\psi\right\|.
\label{eq:dg_weight}
\end{equation} 
For a perfectly symmetrized state $\uline\psi$, we have $w^\Gamma_\mu(\uline\psi)\in \{0,1\}$. 

The starting point for the standard postsymmetrization procedure is to compute the actual set of numbers $w^\Gamma_\mu(\uline\psi_j), j=1\dots d_\mathcal{S}$ for all ($\Gamma,\mu$), which give information on the symmetry content of any state. This information is used to identify subspaces where states of different symmetry have been mixed. 

Then in a second step, symmetrized subspaces for all irreps can be built starting from a suitable set of representatives. Let $k$ be in this set, and let $w^\Gamma_\mu(\uline{\psi_k})$ be larger than the corresponding weight of the other irreps and other partner functions. The projected state $\uline{\overline{\psi}}^\Gamma_{\mu}= P^\Gamma_\mu\uline{\psi_k}/\left\|P^\Gamma_\mu\uline{\psi_k}\right\|$ transforming like $\Gamma,\mu$ is then a good starting point for the symmetrized subspace, the non-projective part of state $k$ is thrown away. For the 1D irreps e.g. $^iE_{3/2}$ this concludes the procedure, but for higher dimensional irreps like $E_{1/2}$ the remaining partner functions must be found to span fully every irrep subspace. One can then use transfer operators $T^\Gamma_{\mu\nu}, \mu\neq\nu$, where $T_{\mu,\nu}^{\Gamma}=\frac{d_{\Gamma}}{|\mathcal{G}|}\sum_{g \in \mathcal{G}}{\left({D^{\Gamma}(g)}_{\mu \nu}^*\vartheta(g)\right)}$ (a generalization of \cref{eq:projector}), and the property $\psi^\Gamma_\mu=T^\Gamma_{\mu,\nu}\psi^\Gamma_\nu$ to generate the basis.

Rather than the standard procedure, we propose below to use a global procedure, which do not throw away any information by selecting representatives. Here, all states will be taken into consideration in one step, rather than operating on every degenerate subspace sequentially. The procedure is not cumbersome, and it is easy to use. It does allow to automatically generate well behaved symmetrized eigenstates of the Hamiltonian.

\subsection{Postsymmetrization using class operators}
\label{sec:ptco}
This procedure is based on the concept of commuting class operators, which are related to conjugation classes. The classes $\mathcal{C}_i$ of a group consist of elements that are conjugate to each other. Two group elements $g_1$ and $g_2$ of the group $\mathcal{G}$ are conjugate if there exists an element $g$ in $\mathcal{G}$ such that $g_2=g(g_1)g^{-1}$. 

Class operators are a set of operators  $\{C_i\}$ constructed by summing operators corresponding to each element of a given class \cite{Chen}
\begin{equation}
C_i=\sum_{g\in\mathcal{C}_i}\vartheta(g).
\label{eq:class_op}
\end{equation}
They are closed under multiplication and commute both with each other and with all group operations. These three properties make them useful e.g. in relation to the sought decomposition of the solution space into orthogonal symmetrized subspaces. 
The number of class operators equals the number of irreps of a group. This yields three classes for the \ctv group. One class consists of the identity, $\mathcal{C}_e=\{e\}$, one class of the three mirror operations, $\mathcal{C}_\sigma=\{\sigma_{vi}, i=\ldots\}$, and the two rotation operations constitute the final class, $\mathcal{C}_3=\{C_3^+,C_3^-\}$. 
The sought symmetrized basis vectors $\uline{\overline\psi}^\Gamma_\mu$ of $\mathcal{S}$ are eigenfunctions of the class operators. The class operator $C_i$ has eigenvalue $\lambda^{\Gamma}_{i}$, i.e. $C_i\uline\psi^\Gamma_\mu=\lambda^{\Gamma}_{i}\uline\psi^\Gamma_\mu$. This eigenvalue can be found directly from the character of the class, $\chi^{\Gamma}_{i}$; one has\cite{Chen}
\begin{equation}
\lambda^{\Gamma}_{i}=\chi^{\Gamma}_{i}\frac{|\mathcal{C}_i|}{d^\Gamma}.
\label{eq:eig_val_co}
\end{equation}

Here $|\mathcal{C}_i|$ is the number of elements in the class $\mathcal{C}_i$. \Cref{eq:eig_val_co} can be used to deduce the group symmetry labels for the eigenfunctions directly without having to employ the projection operator. 

The power of the class operators has been extensively described by \textit{Chen}\cite{Chen}. He showed in particular that the set of class operators linked with a canonical subgroup chain can be used to form a \textit{complete set of commuting operators} (CSCO) for the group space. In our case, since we want a symmetrized basis in the solution space, we must add the Hamiltonian to have a CSCO, since the eigenvalue of the Hamiltonian will be sufficient to distinguish subspaces with identical symmetry. 
For our \ctv case, only one canonical subgroup is relevant, $C_s=\{e,\sigma_{v1}\}$. It will distinguish the partner functions belonging to the 2D irrep $E_{1/2}$. To this end we add the operation $\vartheta(\sigma_{v1})$, which is a class operator for $\mathcal{C}_s$. The CSCO in solution space is then:

\begin{equation}
	CSCO=\{H,{C}_3,{C}_\sigma,\vartheta(\sigma_{v1})\}.
	\label{eq:CSCO}
\end{equation} 
Here the trivial class operator $\mathcal{C}_e$ has been omitted since it corresponds to the identity.

The aim is to find symmetrized states which diagonalize the CSCO. It is then easier to combine all the CSCO operators into one single equivalent CSCO operator $C$ by forming
\begin{equation}
C=\alpha_{H}H+\alpha_3{C}_3+\alpha_\sigma{C}_\sigma+\alpha_s\vartheta(\sigma_{v1})
\label{eq:C_op}
\end{equation}
where the set of factors $\alpha$ must be chosen so that the eigenvalues of $C$ are non-degenerate. In the idealized case in which the states constituting the solution space are perfect eigenstates of $C$, non-degenerate $\alpha$'s would be sufficient. In practice one should however be careful when choosing the factors; guidelines for a good choice will be given in \cref{sec:vb_res}.

Let us now summarize the \tec procedure. From the raw eigenvectors $\{\uline{\psi}_j, j=1...d_\mathcal{S}\}$, a matrix representation is constructed for $C$ using \cref{eq:big_d} and \cref{eq:C_op}. The contributing matrix representation for the Hamiltonian in this raw basis is just a diagonal matrix with the corresponding raw eigenvalues. The unitary matrix $V$, consisting of eigenvectors of $C$ will enable the diagonalization towards the diagonal matrix $\Lambda$

\begin{eqnarray}
V^\dag{C}V=\Lambda. 
\label{eq:diag_O}
\end{eqnarray}

The matrix elements of $V$ provides \textit{in one automated run} the new symmetrized eigenvectors (expressed in the new basis). Identifying the $k$-th symmetrized state as one of the sought symmetrized 
"eigenvectors" $\uline{\overline\psi}^\Gamma_{\mu,l}$, we get $V_{jk}=\braket{\uline\psi_j}{\uline{\overline\psi}^\Gamma_{\mu,l}}$. The overbar stresses that in fact they are not anymore numerically perfect eigenvectors of $H$. Nevertheless, they are in a way more physical since part of their due symmetry, which was broken by numerics, has been restored by the procedure. The index $l$ was introduced to obtain a unique labeling, and is here usually chosen so as to number states with identical symmetry ($\Gamma, \mu$) by order of increasing energy. The symmetrized states can be easily constructed with

\begin{equation}
\uline{\overline\psi}^\Gamma_{\mu,l}=\sum_j\braket{\uline\psi_j}{\uline{\overline\psi}^\Gamma_{\mu,l}}\uline\psi_j.
\label{eq:diag_psi}
\end{equation}

Having obtained a symmetrized basis, we must now identify the three labels $\Gamma, \mu, l$ for all states. The eigenvalues of $C$ can be discarded. It is much more informative to recompute the average values of the original CSCO (\cref{eq:CSCO}), which will provide the labels for the symmetrized states. First,
\begin{equation}
\overline{E}^\Gamma_{\mu,l}=\braket{\uline{\overline\psi}^\Gamma_{\mu,l}}{H\uline{\overline\psi}^\Gamma_{\mu,l}}
\label{eq:corr_energy}
\end{equation}
is nothing else than the new corrected energies. One notes the presence of the index $\mu$: the PTCO cannot restore completely the degeneracies lifted by the original numerical implementation. Nevertheless we still expect closely packed eigenvalues for each $\Gamma$, $l$ subspace. Second, the average values of the \ctv class operators are denoted 

\begin{equation}
\overline{\lambda}^\Gamma_{i,\mu,l}=\braket{\uline{\overline\psi}^\Gamma_{\mu,l}}{C_i\uline{\overline\psi}^\Gamma_{\mu,l}}
\label{eq:class_eigval_id}
\end{equation}
which are closely packed around the ideal value $\lambda^\Gamma_i$ \cref{eq:eig_val_co}. Again there is a slight $\mu$ dependence. By comparison with \cref{eq:eig_val_co} one obtains the characters $\chi^\Gamma_i$ for every state, from which, via the character table, one can deduce the irrep $\Gamma$. Third, the corresponding value of the canonical subgroup class operator $\vartheta(\sigma_{v1})$ will provide the label $\mu$. 

It is trivial to automate also the symmetry recognition step. Alternatively it can be performed using the projection operators \cref{eq:projector}, with the advantage of giving a number which can be interpreted as a degree of symmetry, which is a figure of merit of the achieved symmetrization.

\subsection{A hierarchy of symmetries}\label{sec:hierachy}
With the procedure we have developed, the numerical results will first be symmetrized w.r.t. C$_{\textrm{3v}}$, constructing the CSCO given explicitly in the previous section. The \ctv symmetry is the true symmetry of the structure, only broken by the unsymmetric grid, and numerical errors. From the physical point of view, there might also be higher approximate symmetries like \csv, \dth and \dsh. 

The \csv symmetry group is constructed from \ctv by adding mirror operations w.r.t. three intermediate vertical symmetry planes ($\sigma_{di}, i=1\ldots3$), and \dth is constructed from \ctv by inclusion of the mirror operation of the horizontal symmetry plane ($\sigma_h$). Combining the elements of \csv and \dth leads to the mesoscopic symmetry \dsh.
In both \csv and \dth, there are three double group irreps, $E_{1/2}, E_{3/2}$ and $E_{5/2}$, all two-dimensional. Their identical subduction tables to \ctv yields the correspondence  $E_{3/2}\to^1E_{3/2}\oplus{}^2E_{3/2}$, $E_{5/2}\to E_{1/2}$ and $E_{1/2}\to E_{1/2}$. The subduction is similar for the \dsh group in which there are 6 similar double group irreps of \textit{gerade} and \textit{ungerade} type. The irrep matrices $D^\Gamma$ expanding the set given in Ref.~\cite{dupertuis_msr} are given for all groups in \cref{app:irreps_ctv_dsh}. 

We may investigate the validity of approximate symmetries by attempting to diagonalize a new CSCO constructed with respect to the elevated groups. In such a step, \csv, \dth or \dsh class operators are included in the set of operators to be diagonalized. This implies choosing operators for the $\mathcal{C}_6=\{C_6^+,C_6^-\}$ or the $\mathcal{S}_3=\{S_3^+,S_3^-\}$ classes for \csv and \dth, respectively \cite{Altmann}. These operators enable distinction between $E_{1/2}$ and $E_{5/2}$ irreps. For \dsh we include both of them simultaneously.

One thus sees the versatility of the PTCO, which allows not only to restore the true symmetry, but also to investigate a possible hierarchy of symmetries as ever cruder, but informative, approximations to the problem at hand.

\section{Numerical results}

\label{sec:calc_c3v}
We consider a GaAs QD as described in Sec.\ref{sec:kp_calc}. The length of the hexagon edges is 20 nm and the axial length of the dot is 5 nm. An \algaas shell of thickness 10 nm surrounds the QD in the transverse direction, with infinite potential outside. In the axial direction, the dot is  surrounded by a thick layer of \algaas, sufficient to ensure that all probability distributions goes to zero. The parameters utilized in the \kp Hamiltonian are summarized in Tab.~\ref{tab:Parameters}.
\begin{table}
	\caption{Parameters for \kp simulation, at 0K.}
		\begin{tabular}{ll}
		\hline
	\label{tab:Parameters}
		
Effective el. mass \cite{Vurgaftmanband_param}& \begin{tabular}{l}
												GaAs: $m^*=0.067$\\
												AlAs: $m^*=0.15$ \\\end{tabular}\tb																	
		Luttinger \footnotemark[1] \cite{Vurgaftmanband_param}& \begin{tabular}{llll}
									&$\gamma_1$&$\gamma_2$&$\gamma_3$\\
						GaAs:& 6.9800 &2.0600& 2.9300\\
						AlAs:& 3.7600 &0.8200& 1.4200\end{tabular}\tb
						Bandgap \cite{Vurgaftmanband_param}& \begin{tabular}{l}
												$E_g^{GaAs}=1.519$ eV \\
												$E_g^{AlAs}=3.099$ eV \\																	
		$E_g^{Al_{x}Ga_{1-x}As}$\\$=(1-x)E_g^{GaAs}+xE_g^{AlAs}-x(1-x)c$\\
$c=-0.127+1.310 x$\end{tabular}\tb
\footnotetext[1]{The Luttinger parameters for the \algaas is found using a linear interpolation of the valence band effective masses and the degree of anisotropy, as proposed by Vurgaftman\textit{ et al.}\cite{Vurgaftmanband_param}}
Band offset & \begin{tabular}{l}Ratio: 60.4:39.6 \cite{Yi}\\
$E_v=0.396E_g^{Al_{x}Ga_{1-x}As}$\\
$E_c=0.604E_g^{Al_{x}Ga_{1-x}As}$\\
\end{tabular}\tb 
\hline
		\end{tabular}
\end{table}
\subsection{Numerical implementation, convergence and accuracy}
\begin{figure}[htbp]
	\centering
		{\includegraphics[width=\columnwidth]{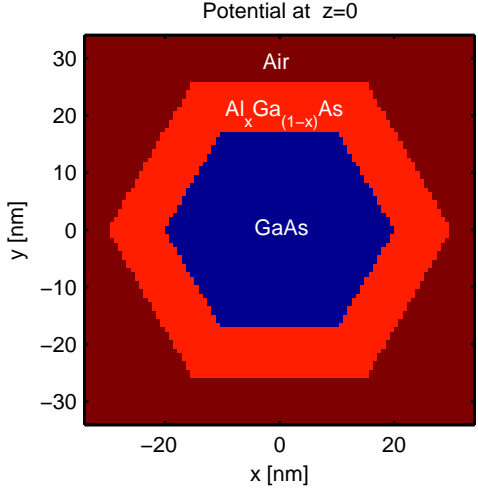}}
	\caption{Cross section of the CB and VB potential in the transverse plane. Note the effect of the square grid on the lateral edges.}
	\label{fig:cs_num_pot}
\end{figure}

The potential is defined on a square grid using 91 points in the $x,y$ and $z$ direction, leading to the cross-section  shown in \cref{fig:cs_num_pot}. On the lateral sides wee see the deviation of the numerical implementation with respect to \csv symmetry. Any effect from this symmetry breaking will later be ideally averaged by the \tec. 

To assess numerical convergence, we define the dimensionless numerical deviation $\Delta_\text{num}$. It is related to $\Delta_E$ which is the energy deviation relative to values obtained at highest resolution:

\begin{equation}
\Delta_\text{num}=\Delta_\text{E}\left(\frac{\hbar^2}{2m_0m^*}\frac{L_\text{cd}}{V_\text{bar}}\right)^{-1}.
\label{eq:deltanum}
\end{equation}
$V_\text{bar}$ is the potential difference between the dot and the first barrier, and $L_\text{cd}$ is a characteristic length of the computational domain, defined as the third root of the computational volume. For the valence band, $m^*$ is replaced by a directionally  averaged  HH mass (\cref{eq:eff_mass}). 
Using $\Delta_\text{num}$ to characterize the convergence allows to extrapolate by comparing the resolution dependence for the conduction and valence band.

\begin{figure}[htbp]
	\centering
		{\includegraphics[width=\columnwidth]{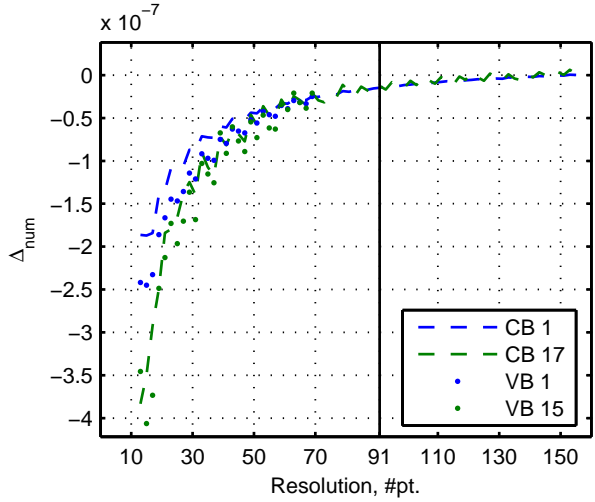}}
	\caption{Numerical deviation $\Delta_\text{num}$ (dimensionless) for conduction band levels number 1 and 17, and for the valence band levels 1 and 15.}
	\label{fig:cb_conv}
\end{figure}

\Cref{fig:cb_conv} shows $\Delta_\text{num}$ for valence band levels 1 and 15, and for conduction band levels 1 and 17 which contains similar dominant envelope functions. Only some low-resolution points are included for the computationally demanding valence band, confirming the convergence dependence. For a resolution of 91 points, we have $\Delta_\text{num}=1.25\cdot10^{-8}$ for conduction band level 17. This corresponds to $\Delta_E=1.7 meV$, and the corresponding extrapolated value for the valence band is -0.3 meV. \Cref{fig:cb_conv} also demonstrates that the relative energy differences within each band are more accurate than the absolute values.

\subsection{Eigenenergies}
\begin{figure}[htbp]
	\centering
		\includegraphics[width=\columnwidth]{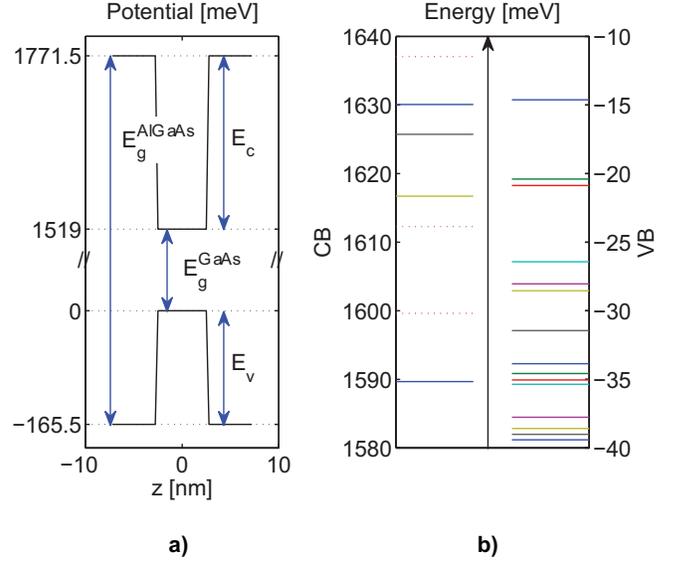}
	\caption{ The potential profile along the axis of the nanowire (a), and the 10 or 15 lowest order energy levels of the conduction and valence band, respectively (b). All levels are twice degenerate due to time reversal. In Fig.~\ref{fig:band_diagram} b), the CB levels marked with a dotted line correspond to $E$ states and are thus four-fold degenerate. Note that the energy scales are not equal.}
	\label{fig:band_diagram}
\end{figure}

The potential profile of the QD is shown in Fig.~\ref{fig:band_diagram}a). In \cref{fig:band_diagram}b) we show the calculated lowest order energy levels for both electrons and holes. All energies are with respect to the top of the valence band. Note that the highest valence band level included here deviates from the top of the valence band by 39meV. In comparison, the split-off band is located a distance $\Delta_\textrm{so}=341$meV from the valence band top.

\subsection{Conduction band eigenstates}
\label{sec:cb_res}
\begin{figure*}
	\centering
		\includegraphics[width=\textwidth]{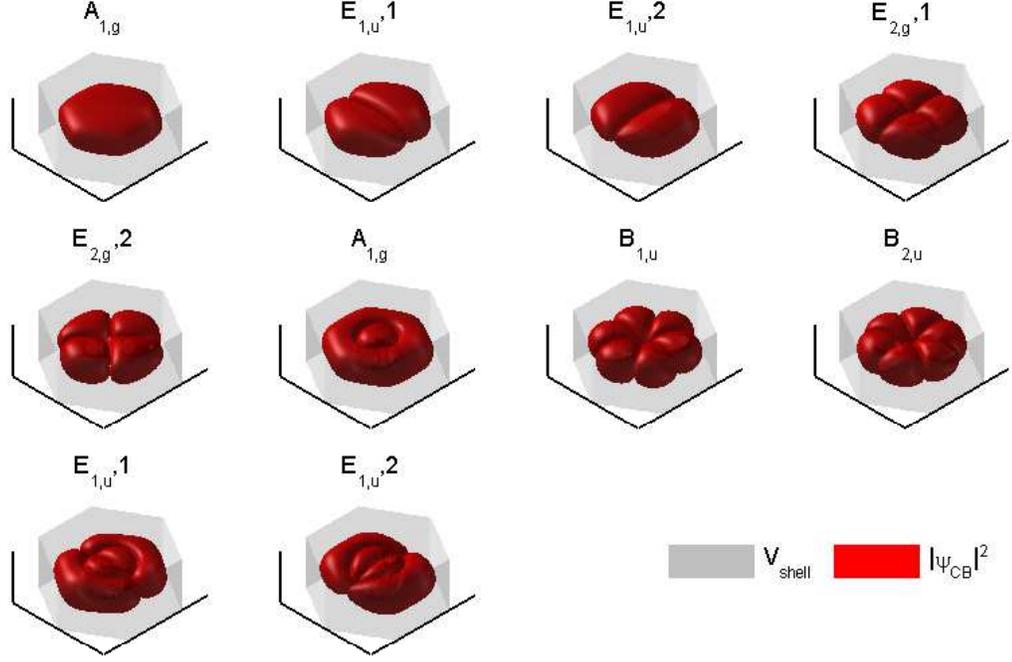}
		\caption{Isosurfaces of probability distributions, $|\psi_\textrm{CB}|^2$, for the conduction band states in order of increasing energy. Each state is twice degenerate due to spin. The barrier potential of the surrounding shell is $V_{\textrm{shell}}$. The \dsh single group symmetry labels are specified for each state. A detailed analysis, summarizing energy and symmetry properties, is given later in \cref{tab:final_class_cb}}.
	\label{fig:cb_3d}
\end{figure*}

Probability distributions $|\psi_n|^2$ for the 10 lowest levels in the conduction band are shown in Fig.~\ref{fig:cb_3d}. The full spinorial eigenfunctions $\uline{\psi_n}$ can be constructed from the 
scalar envelope functions $\psi_n$ \cite{dupertuis_opt}

Since the states of the conduction band have the full \dsh symmetry in the present model, we label every envelope function in Fig.~\ref{fig:cb_3d} by \dsh single group irreps. The \dsh irreps are labeled with subscripts $g$ and $u$, depending on whether they are even (\textit{gerade}) or odd (\textit{ungerade}) with respect to inversion. The gerade single group irreps are the 1D irreps $A_{1,g},A_{2,g},B_{1,g},B_{2,g}$, and the 2D irreps  $E_{1,g}$ and $E_{2,g}$, similarly for the ungerade irreps($g\to u$). A complete summary of energies and symmetry labels is given later, in \cref{tab:final_class_cb}.

\subsection{Valence band eigenstates}
\label{sec:vb_res}
In the analysis of the valence band states, we will utilize extensively the \tec outlined in \cref{sec:postsymmetrization}. First one needs to symmetrize the eigenstates w.r.t. the true \ctv symmetry of the original problem, as the small lack of symmetry due to the grid  may strongly influence the choice of eigenstates within degenerate subspaces. The \tec will also give quantitative information about the effect of the symmetrization process. In a second step we investigate approximate elevated symmetries with the same technique.

%\subsubsection{Analysis of true \texorpdfstring{\ctv}{c3v} symmetry}
\subsubsection{Analysis of true \ctv symmetry}
The combined CSCO operator, $C$, used for postsymmetrization according to \ctv is given by \cref{eq:C_op}; the $\alpha$ parameters were left free in \cref{sec:ptco}. From the numerical point of view, however, they must be chosen carefully to ensure well separated eigenvalues. To this end let us consider the eigenvalues of $C$. It is easy to show from the eigenvalue equation
\begin{eqnarray}
C\overline{\uline\psi}^\Gamma_{\mu,l}&=&c^\Gamma_{\mu,l}\overline{\uline\psi}^\Gamma_{\mu,l}
\label{eq:csco_eig_val_gen}
\end{eqnarray}
that one has
\begin{equation}
c^\Gamma_{\mu,l}=\alpha_H\overline{E}^\Gamma_{\mu,l}+\sum_{i=3,\sigma,s}\alpha_i\overline\lambda^\Gamma_{i,\mu,l}
\label{eq:csco_eig_val}
\end{equation}
where $\overline{E}^\Gamma_{\mu,l}$ and $\overline\lambda^\Gamma_{i,\mu,l}$ are defined by \cref{eq:corr_energy,eq:class_eigval_id}. The $c^\Gamma_{\mu,l}$ eigenvalues can be chosen freely, as they are directly related to the $\alpha_k$ coefficients.  Since we know the ideal analytical eigenvalues $\lambda^\Gamma_i$ of the class operators \cref{eq:eig_val_co}, and that the corrected energies $\overline{E}^\Gamma_{\mu,l}$ will be close to the original uncorrected energies it becomes possible to choose the $\alpha$ coefficients \textit{a priori} in such a way that the CSCO eigenvalues  $c^\Gamma_{\mu,l}$ will be well separated. To this end we consider the approximate spectrum of \cref{eq:csco_eig_val}:
\begin{eqnarray}
\tilde{c}^\Gamma_{\mu,l}&=&\alpha_H\overline{E}^\Gamma_{\mu,l}+\sum_{i=3,\sigma}
\alpha_i\lambda^\Gamma_{i}+\alpha_s\lambda_{s,\mu}
\label{eq:csco_eig_val_approx}
\end{eqnarray}
Note that we singled out the last parameter $\alpha_s$ which is related to the  class operator of the subgroup $\mathcal{C}_s$ distinguishing different partner function indices $\mu$: to avoid confusion between $\lambda^\Gamma_i$ of \ctv and $\mathcal{C}_s$ we denoted the latter eigenvalue $\lambda_{s,\mu}$.

Since one of the eigenvalues may be chosen arbitrarily, it is convenient to shift $C$ by $\alpha_HE_0$, where $E_0$ is the ground state energy, and to make $c^\Gamma_{\mu,l}$ dimensionless by setting $\alpha_{H}=1/\max_{\Gamma,\mu,l}(\overline{E}^\Gamma_{\mu,l}-E_0)$. One then obtains the set of eigenvalues
\begin{equation}
c^\Gamma_{\mu,l}=\frac{\overline{E}^\Gamma_{\mu,l}-E_0}{\displaystyle\max_{\Gamma,\mu,l}{\overline{E}^\Gamma_{\mu,l}-E_0}}+\tau^\Gamma_{\mu}
\label{eq:csco_eig_val_dim_less}
\end{equation}
such that the range of the energy splitting is normalized to unity. Here
\begin{equation}
\tau^\Gamma_{\mu}=\alpha_s\lambda_{s,\mu}+\sum_{i=3,\sigma}\alpha_i\lambda^\Gamma_{i}.
\label{eq:alpha_lambda}
\end{equation}
 
The explicit choice of $\tau^\Gamma_{\mu}$ is given in \cref{tab:eig_val_big_d}, the $\alpha$'s can be easily obtained by inverting \cref{eq:alpha_lambda}

Several considerations have been taken into account in our choice of $\tau^\Gamma_\mu$. 
In particular their relative magnitude decides which aspects are given highest priority. E.g. choosing $\tau^\Gamma_{\mu}\gg1$ would emphasize very large separation w.r.t. group symmetry labels and enforce symmetry at the price of eventually remixing completely the Hamiltonian eigenstates. On the opposite, $\tau^\Gamma_{\mu}\ll1$ would give negligible weight to symmetry considerations. We have chosen an intermediate set of values, where the relative step between neighboring energy levels are a bit smaller than the difference in eigenvalue for two functions belonging to different irreps. 
This choice was found appropriate for our structure, where the symmetry of degenerate states was slightly broken by a non-symmetric grid.
\begin{table}
\begin{tabular}{cccccc}
&$\tau^{^{1}E_{3/2}}$&${\tau^{^{2}E_{3/2}}}$&$\tau^{E_{1/2,1}}$&$\tau^{E_{1/2,2}}$\\
&-3.5&-1.5&0.5&2.5\\
\end{tabular}
\caption{Parameters $\tau^\Gamma_{\mu}$ chosen to separate between \ctv irreps and partner functions}
\label{tab:eig_val_big_d}
\end{table}

\begin{figure}[htbp]
	\centering
		\includegraphics[width=\columnwidth]{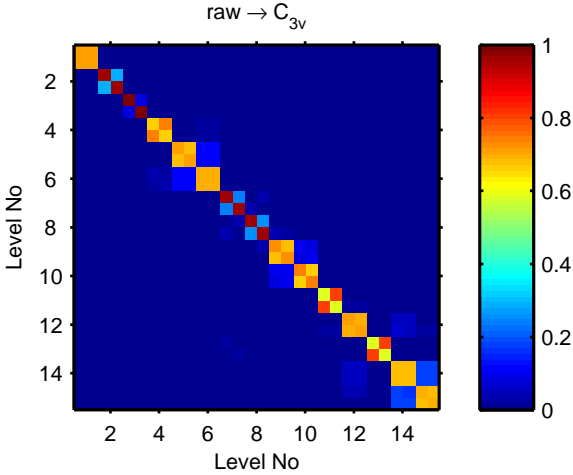}
		\caption{Norms of the matrix elements of the unitary transformation symmetrizing the eigenstates w.r.t. \ctv.
	\label{fig:unitary_c3v}}
\end{figure}

\begin{figure}[htbp]
	\centering
		\includegraphics[width=\columnwidth]{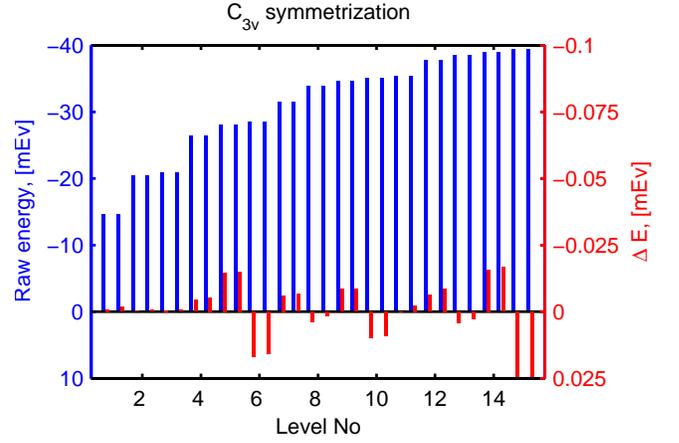}
		\caption{Changes in energies due to the symmetrization w.r.t. \ctv. The raw energy spectrum is shown in blue, and the difference between the symmetrized and the raw spectrum is shown in red.}
	\label{fig:en_diff}	
\end{figure}

Having now determined the set $\alpha_k$, we computed the $C$ operator in a raw basis comprising the first 30 valence band states (including Kramers degeneracy) ordered by energy, and diagonalized it. To visualize the effect of mixing due to eigenstate symmetrization we plot in \cref{fig:unitary_c3v} the norm of the matrix elements of the unitary transformation matrix (\cref{eq:diag_psi}). \Cref{fig:en_diff} display the change in energy $\Delta$E after symmetrization, confirming that the changes are modest. %One sees clearly that the most important corrections are when different Kramers doublets of very different energies are  mixed. The energy changes remain however modest. 

At large, the transformation matrix remixes states only within  degenerate 2x2 subspaces corresponding to distinct Kramers doublets. There are no intertwined blocks, but the procedure would work even in this case, sorting correctly the partner functions and symmetrize them. In the raw states there were also some mixing between certain neighboring energy levels, as evidenced by the sequence of energies shown in \cref{fig:en_diff}. In this finer level one sees more subtle mixing extending over different Kramers doublets e.g. levels 5 and 6. This is a result of the nonsymmetric grid,  and would be cumbersome to correct using a stepwise projection operator procedure rather than the automated \tec. Such mixing increase for higher excited states, as expected. These include higher spatial frequencies, and will accordingly  be more strongly influenced by limited grid resolution. 

\begin{figure}[htbp]
	\centering
		\includegraphics[width=\columnwidth]{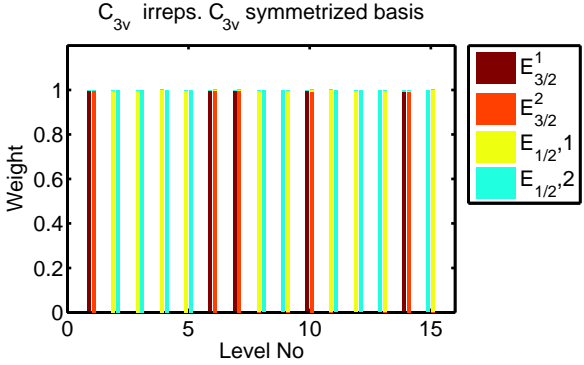}
		\caption{Double group weights $w^\Gamma_\mu\left(\overline{\uline\psi}^\Gamma_\mu\right)$ w.r.t. the irreps of the \ctv group after \ctv postsymmetrization}
	\label{fig:dg_sym_c3v_c3v_c3v}
\end{figure}

After symmetrization, we also analyze the \ctv irreps associated with different spinors. Fig.~\ref{fig:dg_sym_c3v_c3v_c3v}. shows the double group weight for the 30 lowest order states, divided again into 15 Kramers degenerate doublets. The eigenstates are very purely symmetrized w.r.t. \ctv ($w^\Gamma_\mu\left(\overline{\uline\psi}^\Gamma_\mu\right)\approx 1$), confirming the symmetry of the QD, and the postsymmetrization procedure. The small deviation for the highest order states is again due to grid inaccuracies. From \cref{fig:unitary_c3v,fig:en_diff,fig:dg_sym_c3v_c3v_c3v}, we see that all the states that were most strongly symmetry mixed in the raw eigenspace did correspond to Kramers doublets with very small energy separation. However, there were also other off-diagonal mixing leading to the largest changes in energy.

\begin{figure}[htbp]
	\centering
		\includegraphics[width=\columnwidth]{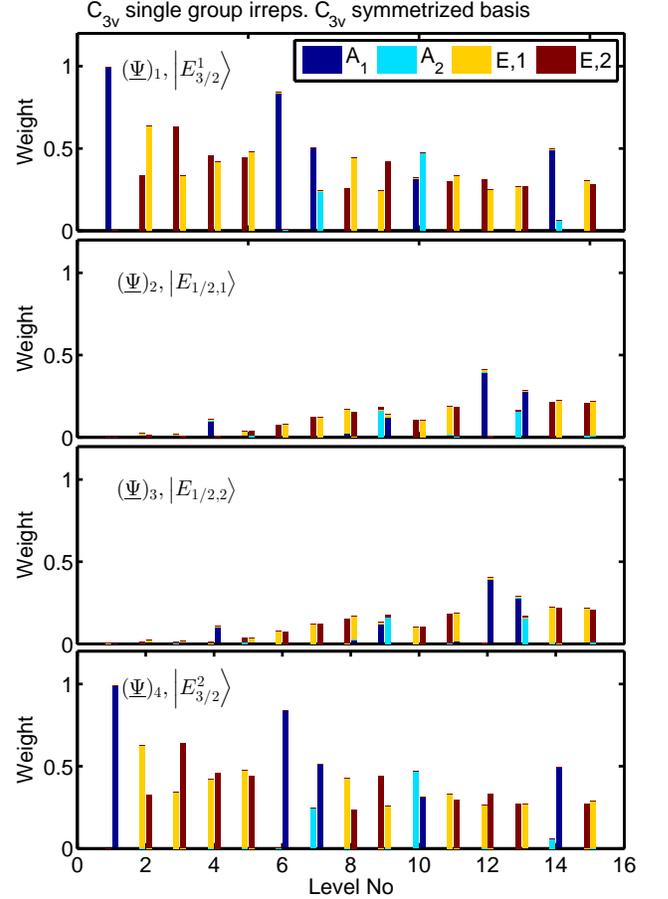}
	\caption{Single group weight $w^\Gamma_\mu\left(\overline{\psi}^\Gamma_\mu\right)$ with $\Gamma$, $\mu$ being single group irreps, for each spinor component in the HSBF basis after \ctv postsymmetrization.}
	\label{fig:sg_sym_c3v}
\end{figure}

We proceed to the analysis w.r.t. UREFs and single group irreps, which is shown in \cref{fig:sg_sym_c3v}. The weights $w^\Gamma_\mu(\psi_i)$, plotted for each HSFB component, $\psi_i, i=1\ldots4$ must be distinguished from the UREF weight $w^{\Gamma_b,\Gamma_a}(\uline\psi^\Gamma_\mu)$ defined in \cref{eq:weight_uref}. Using \cref{eq:weight_uref,eq:dg_weight} and Eq.~(48) of Ref.~\citenum{dupertuis_msr}, it is easy to see that they differ only by a  Clebsch-Gordon coefficient:\gbt{check squared}
\begin{equation}
w^{\Gamma_a}_{\alpha}(\psi^{\Gamma,\Gamma_b}_{\mu,\beta})=\left|C^{\Gamma,\Gamma_b^*;\Gamma_a}_{\mu,\beta;\alpha}\right|w^{\Gamma_b,\Gamma_a}(\uline\psi^\Gamma_\mu)
\end{equation}  
We use  $w^\Gamma_\mu(\psi_i)$  to be able to check the content of every HSBF component, in particular when it vanishes or when the weight is supposed to be equally distributed onto separate components. $w^\Gamma_\mu(\psi_i)$ has easy interpretation, yielding  directly the fractions of one given spinor component transforming according to the single group irreps. For the \ctv case, the two definitions differ by a factor $\sqrt{2}$ for the weight in the central components of the $E_{1/2}$ spinors, but the definitions are equivalent for the remaining UREFs.

Considering \cref{fig:sg_sym_c3v}, we first note that the UREFs are in accordance with \cref{eq:vb_spinors}. However, there is a striking asymmetry between weights of partner functions 1 and 2 in the $\ket{^iE_{3/2}}$ components of the $E_{1/2}$ states, most notably for levels 2 and 3 and 8 and 9. Due to time reversal symmetry, the two $\ket{^1E_{3/2}}$ components  and the two $\ket{^2E_{3/2}}$ components, should correspond to partner functions of $E$ with the \textit{same weights}. Such asymmetry can occur due to mixing of states from nearby energy levels if the states bear the same symmetry label. In 
this case, they are mixed in a way that cannot be sufficiently corrected  by the \ctv symmetrization procedure. This explanation is supported by the fact that neighboring $E_{1/2}$ levels have opposite weight imbalance in \cref{fig:sg_sym_c3v}, and also by their nearby energies  in \cref{fig:en_diff}. We interpret such mixing as being a result of the imperfect grid and quasi-degeneracies. However, the presence of additional quasi-degeneracies should alert us of the possible existence of an approximate elevated symmetries, possibly able to correctly remix the relevant states.

%\subsubsection{Analysis of elevated \texorpdfstring{\csv}{csv} and \texorpdfstring{\dth}{dth} symmetries}
\subsubsection{Analysis of elevated \csv and \dth symmetries}
\begin{figure}[htbp]
	\centering
		\includegraphics[width=\columnwidth]{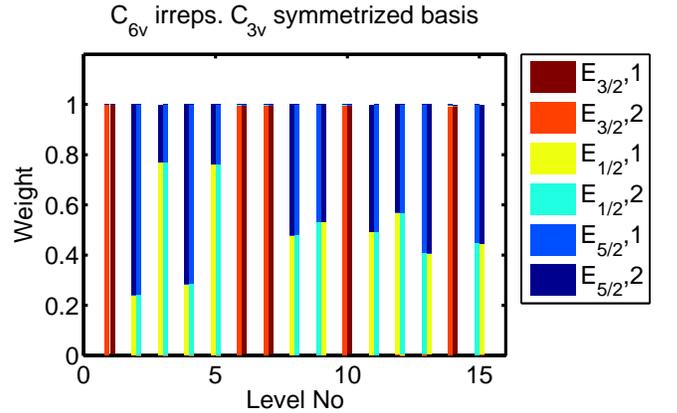}
		\caption{Double group weights w.r.t. the irreps of the elevated \csv group. The eigenstates have been symmetrized w.r.t. the \ctv group.}
	\label{fig:dg_sym_c3v_c3v_c6v}
\end{figure}

The quasi-degeneracy of the first two $E_{1/2}$ levels hinted  at the possible occurrence of approximate elevated symmetries. Such symmetries have previously been found both in theoretical \cite{dupertuis_opt} and experimental \cite{karlsson,Dupertuis_letter} work on [111] oriented \ctv heterostructures. Because the mesoscopic structure has \dsh symmetry, we shall test the possible existence of \csv, \dth and \dsh elevated symmetries to see if there is a hierarchy and, if so, how it is organized.

Fig.~\ref{fig:dg_sym_c3v_c3v_c6v} show the weights of the \ctv symmetrized states according to irreps of the \csv symmetry group. Very similar results were obtained for \dth. We see no change for $^iE_{3/2}$ states which are also well defined using these elevated symmetry groups. This is in agreement with the subduction table for \csv to \ctv which yields $E_{3/2}{\to}E_{3/2}^1{\oplus}E_{3/2}^2$, the same applies for \dth. 
For the remaining irreps, the corresponding subduction yield $E_{5/2}{\to}E_{1/2}$ and $E_{1/2}{\to}E_{5/2}$. This is in accordance to the results of  \cref{fig:dg_sym_c3v_c3v_c3v,fig:dg_sym_c3v_c3v_c6v}. However, we see clearly some irrep mixing which prevents unique labeling of these states in view of the elevated symmetry groups \csv and \dth.

For closer investigation we shall perform a new symmetrization, as proposed in \cref{sec:postsymmetrization}.
To this end, one extends the set of operators in the CSCO with class operators for classes of \csv or \dth. This can be easily done by adding the operators of $\mathcal{C}_6$ or $\mathcal{S}_3$, respectively. There is no need to add further class operators for canonical subgroups here. This step allows to distinguish between the $E_{1/2}$ and $E_{5/2}$ irreps of the elevated groups. Of course one cannot expect the output of these elevated symmetrizations to be exact since they are not true symmetries of the QD. Our choice of symmetrization parameters is given in \cref{tab:eig_val_big_d_elev}.

The parameters were again chosen to separate clearly the different irreps. Note however that the separation between $\tau^\Gamma_\mu$ for irreps that are indistinguishable in \ctv has been chosen smaller than the separation between those corresponding to different \ctv irreps, and also smaller than 1, giving the Hamiltonian priority over the approximate symmetry.
\begin{table}
\begin{tabular}{cccccccc}
&$\tau^{E_{3/2}}_1$&${\tau^{E_{3/2}}_2}$&$\tau^{E_{1/2}}_1$&$\tau^{E_{1/2}}_2$&$\tau^{E_{5/2}}_1$&$\tau^{E_{5/2}}_2$\\
&-3.5&-1.5&0.65&2.65&0.35&2.35\\

\end{tabular}
\caption{Parameters $\tau^\Gamma_\mu$ for symmetrization w.r.t. elevated symmetry groups \ctv and \dth.}
\label{tab:eig_val_big_d_elev}
\end{table}

\begin{figure}[htbp]
	\centering
		\includegraphics[width=\columnwidth]{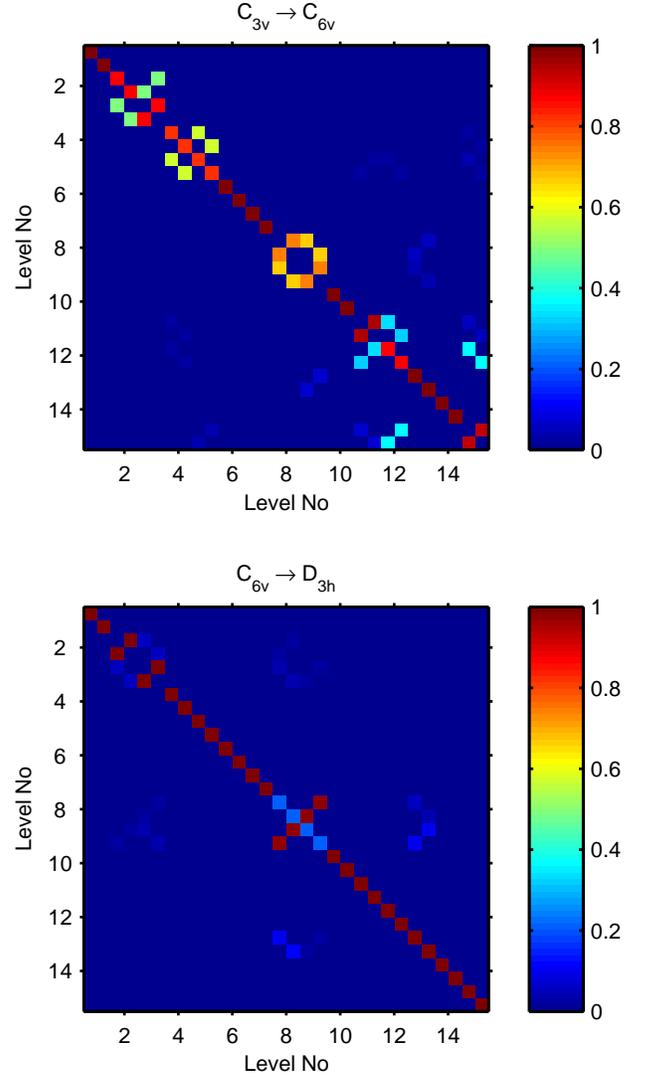}
		\caption{Norms of the matrix elements of $V_2$ and $V_3$, connecting the \csv-symmetrized and \dth-symmetrized states to \ctv- and \csv- symmetrized bases respectively.}
	\label{fig:unitary_c3v_to_elevated}
\end{figure}

We refer to the new eigenstates as \csv- and \dth- symmetrized, and denote the unitary transformation matrices from \ctv to \csv by $V_2$, and from \csv to \dth by $V_3$. The norms of their matrix elements are shown in \cref{fig:unitary_c3v_to_elevated}.

The transition from \ctv to \csv (\cref{fig:unitary_c3v_to_elevated} a)) leaves the $^iE_{3/2}$ states practically unaltered, as expected from 
\cref{fig:dg_sym_c3v_c3v_c6v}. These states were already well symmetrized by the basic \ctv symmetrization step. As expected we also see that quasi-degenerate subspaces that had 
$E_{1/2}$ symmetry in \ctv are now remixed, namely levels 2 and 3, 4 and 5  and also 8 and 9. Furthermore we see in the lower right corner of \cref{fig:unitary_c3v_to_elevated} 
a), three non-diagonal blocks mixing levels 11, 12 and 15 non-trivially together.  The asymmetric structure of the intermixing blocks might indicate that a level above level 15 
should also have been included in the solution space. Block structures like this are useful in revealing nearby states of similar symmetry. 

Interestingly, symmetrization w.r.t. \csv and \dth have practically the same effect on the eigenstates. This is clearly seen in \cref{fig:unitary_c3v_to_elevated} b), showing the structure of the unitary transformation, $V_3$, between these two eigenstate bases. The main effect is merely an irrelevant reordering of some neighboring states. For this reason we shall not any more comment \dth separately. Indeed the similarity of \csv and \dth indicate that \dsh symmetry will be the most relevant elevated symmetry.

\begin{figure}[htbp]
	\centering
		\includegraphics[width=\columnwidth]{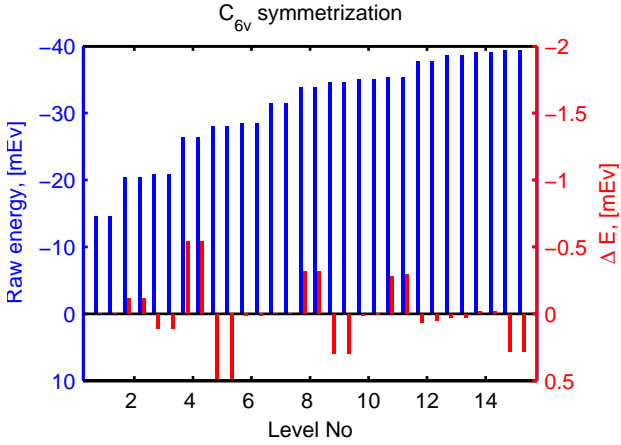}
		\caption{Changes in energies due to the symmetrization w.r.t. \csv (very similar results hold for \dsh symmetrization). The raw energy spectrum is shown in blue, and the difference between the symmetrized and the raw spectrum is shown in red.}
	\label{fig:en_diff_c6v}	
\end{figure}

It has still not been demonstrated that the elevated symmetry \csv is in fact a good approximation to the true solution of the original Schrödinger equation. In \cref{fig:en_diff_c6v}, we show the variation of the energy levels due to the imposed symmetry \csv. One sees that the changes in energy remain very small ($\Delta E< 0.6$ meV) compared to the confinement energies, confirming the continued validity of the Schrödinger equation. Not surprisingly, the alterations brought on by the elevated \csv symmetrization does however lead to $\Delta E$ an order of magnitude bigger than for the \ctv symmetrization (\cref{fig:en_diff}). Clearly this is a manifestation that symmetries that are not true symmetries of the physical system \textit{are imposed}. The \csv symmetrization tend to bring closer energy levels with intermixed states. We also see that $\Delta E$ is biggest when non-degenerate subspaces are remixed, as expected.

\begin{figure}[htbp]
	\centering
		\includegraphics[width=\columnwidth]{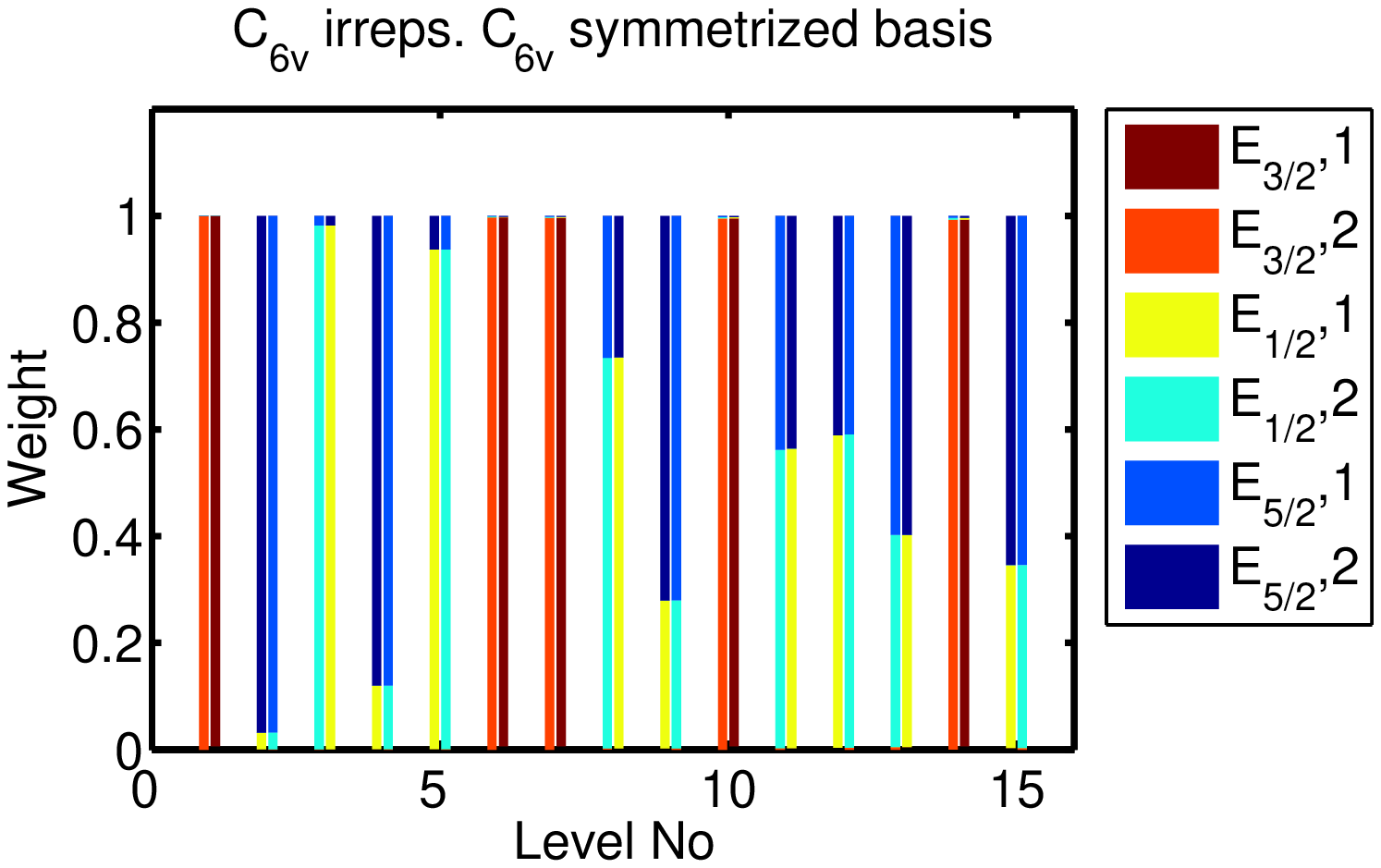}
		\caption{Double group weights $w^\Gamma_\mu(\overline{\uline{\psi}}^\Gamma_\mu)$ of the irreps of the \csv group after \csv symmetrization of the basis.}
	\label{fig:dg_sym_c3v_c6v}
	\end{figure}

The weights of the \csv double group irreps for \csv-symmetrized eigenstates are shown in \cref{fig:dg_sym_c3v_c6v}. The identification of the \csv double group irreps subduces correctly towards the previous identification with \ctv (\cref{fig:dg_sym_c3v_c3v_c3v}). The most interesting aspect of \cref{fig:dg_sym_c3v_c3v_c6v} is certainly that the first two levels which were characterized by $E_{1/2}$ \ctv irreps (levels 2 and 3) are now distinct \csv irreps, with nearly pure character, similarly for levels 4 and 5. As the two latter levels were not as much degenerate, their energy change is much larger (\cref{fig:en_diff}). The same comment also applies to levels 8 and 9, although we see that these levels do \textit{not} seem to be well-described by the elevated symmetry. This is not surprising since states of higher energy levels are generally more sensitive to symmetry breaking contributions due to a microscopic structure. For the lower levels, the remixed states are well defined by $E_{1/2}$ or $E_{5/2}$ irrep labels, accordingly they approximately obey the symmetry of the elevated symmetry groups.  

\begin{figure}[htbp]
	\centering
		\includegraphics[width=\columnwidth]{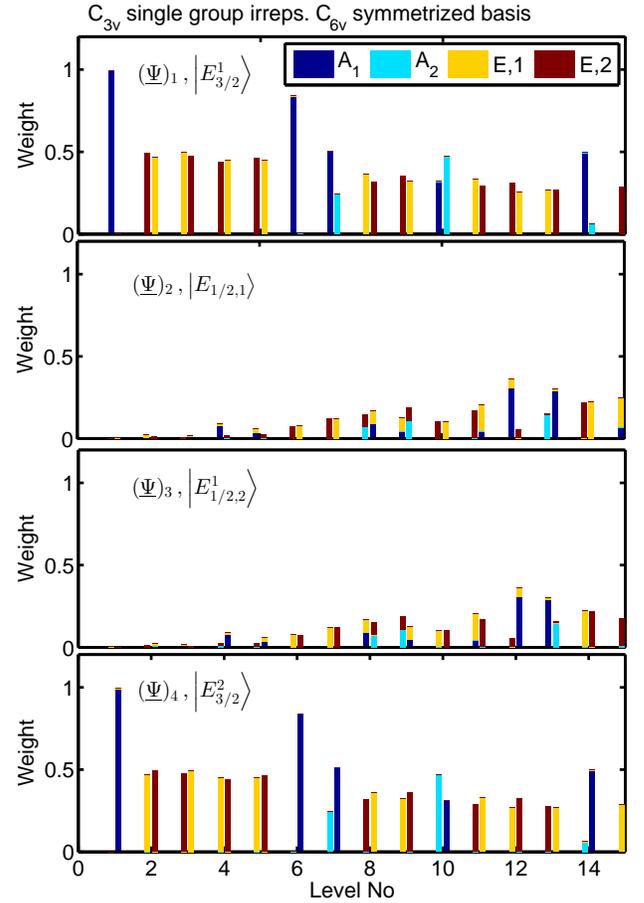}
	\caption{Weights $w^\Gamma_\mu(\overline{\psi}^\Gamma_\mu)$ for each component of the HSBF basis, where $\Gamma$ $\mu$ are single group irreps and partner functions of the \ctv group. The eigenstates have been transformed using the \csv-symmetrization.}
	\label{fig:sg_sym_c3v_c6v}
	\end{figure}

The symmetries of the spinor components of the \csv symmetrized eigenstates are given in \cref{fig:sg_sym_c3v_c6v} (using only \ctv irreps, for easier comparison with \cref{fig:sg_sym_c3v}). The single group decomposition of the components of the $^iE_{3/2}$ states obviously remains the same as in \cref{fig:sg_sym_c3v}. It is more interesting  to consider the $E_{1/2}$ states, especially those corresponding to levels 2 and 3, and also the states of levels 8 and 9. For these there is a meaningful difference. The weight imbalance present in \cref{fig:sg_sym_c3v} has essentially disappeared in the new eigenstate basis (\cref{fig:sg_sym_c3v_c6v}). This is a proof that  there was indeed a mixing between the quasi-degenerate levels due to  grid imperfections that could not be retrieved by the \ctv symmetrization, since the involved states bared the same labels! In the elevated \csv group, since mixed states bear different irreps they are easily disentangled. The restoration of the correct balance between the weights of	partner UREFs bearing \ctv single group labels is also a clear signature of this fact. 

\subsubsection{\dsh as the ultimate elevated symmetry ?}
The results from the symmetrization analysis in view of the elevated symmetries \dth and \csv revealed a very close relation between these symmetry elevations. This is a clear indication that the relevant elevated approximate symmetry group may in fact be \dsh which collects all the symmetry elements of \dth and \csv. In fact this is not entirely surprising as the heterostructure symmetry is \dsh, only broken at the microscopic level by the crystal symmetry which in principle does lower it to \ctv. This motivates a symmetrization w.r.t. the \dsh symmetry, i.e. with inclusion of class operators for both $\mathcal{C}_6$ and $\mathcal{S}_3$ in the CSCO. The double group irreps of \dsh carry the same types of labels as  \csv and \dth, however there are twice as many as one has gerade and ungerade kinds. The $\tau^\Gamma_\mu$ parameters  for the \dsh irreps are given in \cref{tab:eig_val_big_d_dsh}, the values for the gerade/ungerade were obtained by adding/subtracting 0.1 to the corresponding values in \cref{tab:eig_val_big_d_elev}.

\begin{table}
\begin{tabular}{cccccccc}
&$\tau^{E_{3/2,i}}_1$&${\tau^{E_{3/2,i}}_2}$&$\tau^{E_{1/2,i}}_1$&$\tau^{E_{1/2,i}}_2$&$\tau^{E_{5/2,i}}_1$&$\tau^{E_{5/2,i}}_2$\\
$i=g$&-3.6&-1.6&0.75&2.75&0.45&2.45\\
$i=u$&-3.4&-1.4&0.55&2.55&0.25&2.25
\end{tabular}
\caption{Parameters $\tau^\Gamma_\mu$ for symmetrization w.r.t. elevated symmetry group \dsh.}
\label{tab:eig_val_big_d_dsh}
\end{table}

\begin{figure}[htbp]
	\centering
		\includegraphics[width=\columnwidth]{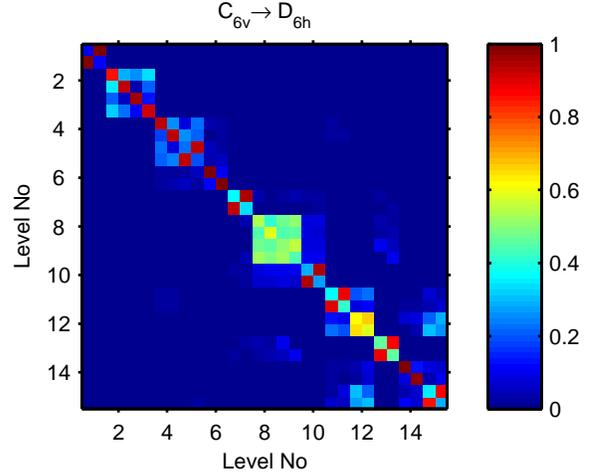}
		\caption{Unitary transformation between bases symmetrized w.r.t. \csv and \dsh symmetries.}
	\label{fig:unitary_c6v_dsh}
\end{figure}

In \cref{fig:unitary_c6v_dsh} we first display the norm of the matrix elements of the unitary transformation bringing the \csv-symmetrized basis to the \dsh symmetrized basis.  Except for levels 8, 9, 11, 12 and 15 there are little changes apart from trivial reordering of states. The remixing of the highest excited states cannot be analyzed due to the stringent limitation in the number of states kept in the solution space, but the remixing of levels 8 and 9 can be clearly interpreted as a further attempt to give distinct irreps $E_{3/2,u}$ and $E_{5/2,u}$ to these levels, that were  not completely  separated in the \csv symmetrization ($E_{1/2}$ and $E_{1/2}$ in \cref{fig:dg_sym_c3v_c3v_c6v}). We clearly see that the departure from \dsh symmetry, due to the underlying symmetry of the Zinc Blende lattice, starts to be important after level 10, in particular for levels 11,12 and 15 which have strongly mixed character with respect to \dsh symmetry.

\begin{figure}[htbp]
	\centering
		\includegraphics[width=\columnwidth]{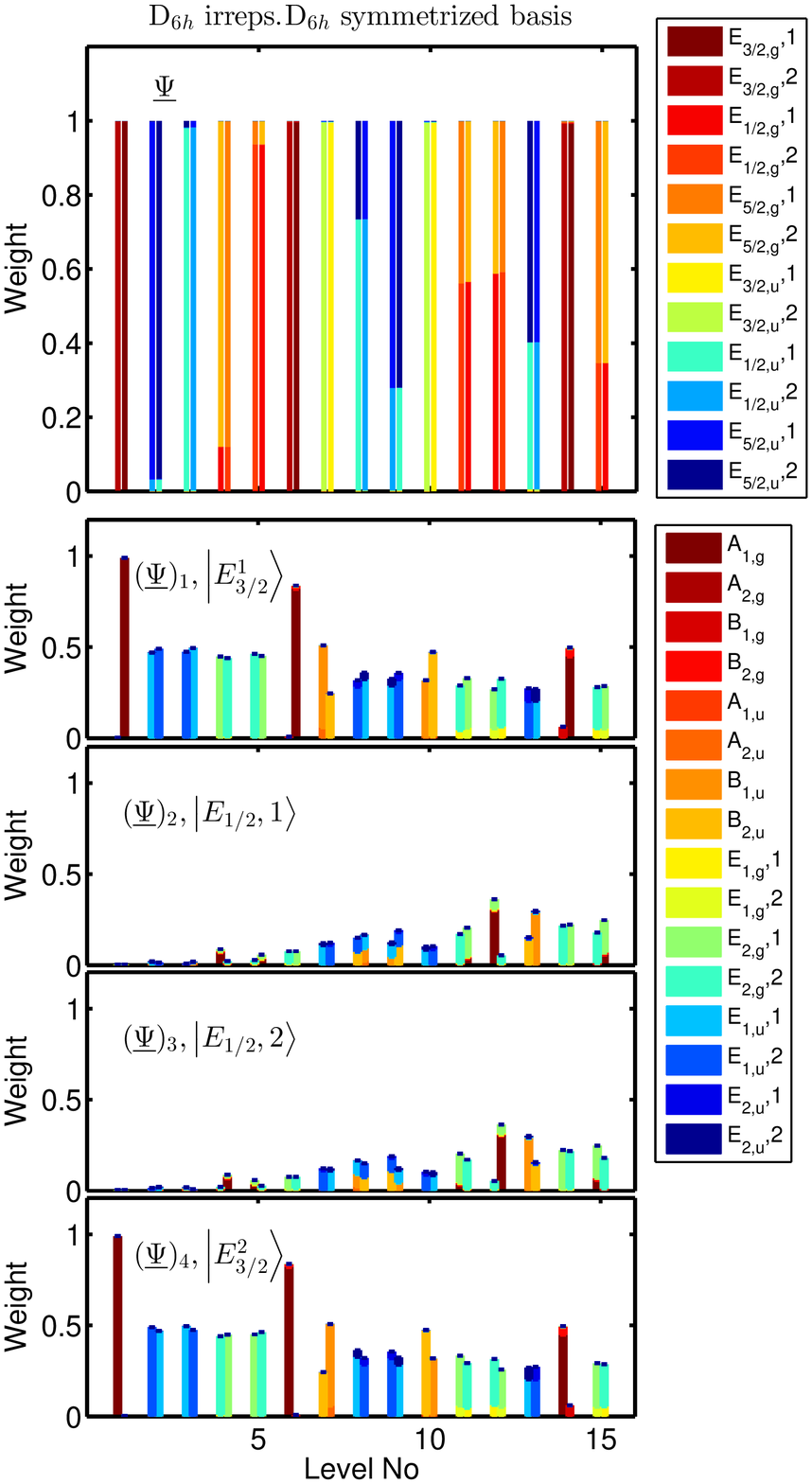}
		\caption{Double group weights and single group weights for each HSBF component  w.r.t. \dsh group,  using the \dsh symmetrized basis.}
	\label{fig:d6h}
\end{figure}

The energy changes $\Delta E$ due to the \dsh symmetrization are given in \cref{tab:final_class}. From this we see again that neighboring levels bearing $E_{1/2}$ labels in view of \ctv are renormalized. The energy differences are very similar to the corresponding values for \csv (\cref{fig:en_diff_c6v}). In \cref{fig:d6h}a), we show the double group weights of the \dsh-symmetrized eigenstates. The purities of the states w.r.t. the double group irreps displayed in  \cref{fig:d6h}a) are also very similar to the corresponding \csv analysis (\cref{fig:dg_sym_c3v_c3v_c6v}). The \dsh symmetry is thus fulfilled to approximately the same degree as the intermediate symmetries \csv and \dth, as is also confirmed by the deviation in energy due to the \dsh symmetrization (\cref{tab:final_class}), which stays on the same order as for \csv. 

The single group analysis of each spinorial component in the HSBF basis in \cref{fig:d6h} b) leads to results in fair agreement with the UREFs for \dsh which are listed in \cref{app:uref_dsh}. 
These results fully confirm the relevance of the approximate symmetry elevation to \dsh for the valence band in our structure, and clearly shows when it applies and when it has limitations. In the following section a more in depth analysis of the UREF weights of \cref{fig:d6h} is carried out.

\subsection{Characterization of DPGPS, UREF's and envelope function symmetry of valence band eigenstates}
Up to now we have concentrated on the symmetry properties of eigenstates as a whole. It is however possible to get important information and insight within each of them by studying the nature of the UREFs, identifying dominant HSBF (DPGPS) and dominating UREF. Such an approach will not only provide enhanced physical interpretation and intuition, but can also reveal to be extremely helpful information when building more complex objects, e.g. multiexcitons in a configuration interaction approach and in the strong confinement limit (c.f. \citenum{Dupertuis_letter, karlsson}).

The analysis will be carried out using essentially elevated \dsh symmetry which provides the finest classification, establishing the DPGPS in \cref{sec:dpgps} and determining the dominant UREF's in \cref{sec:urefs}. The analysis is finally completed in \cref{sec:spatial} by an analysis in terms of more classical azimuthal and radial quantum numbers, %which provides intuitive understanding for the sequence of irreps and outlines the key role of symmetry adapted functions for making the link. 
which provides an intuitive understanding for the sequence of irreps and outlines a link to symmetry adapted functions \cite{Altmann}. 

The main results of this section are summarized in \cref{tab:final_class}. Together with \cref{fig:d6h}, \cref{tab:final_class} provides a deep understanding of all valence band states which will allow to decipher the remarkable properties of the optical transitions.

\subsubsection{Heavy and Light hole mixing, HSBFs and DPGPS}
\label{sec:dpgps}

The most common way to analyze valence band states in heterostructures is in terms of valence band mixing between HH and LH states. This is an import from quantum well physics where, at zone-center, there is no mixing. In strongly oblate (c.f. \cref{fig:structure}) disk-like structures this is still a good starting point due to weak mixing for the ground states, allowing to deduce that the ground state should be HH-like. Moreover one can predict further the order of excited HH- and LH-like states coarsely using a scalar anisotropic effective mass model based on Eqs.~\ref{eq:eff_mass} which neglects valence band mixing and assumes an infinite cylindrical potential well with the same height and cross sectional area as the nanowire QD. One obtains a separation of about 45 meV between the ground HH and LH states. Accordingly, we do expect the 15 lowest energy levels (\cref{fig:band_diagram}) to be dominated by HH states, in agreement with what can be seen in \cref{fig:sg_sym_c3v,fig:d6h}, where it is easy to see that the HH-weight always remains above 0.5 using \cref{eq:lh_weight,eq:lh_weight_hsbf}. We also see clearly in \cref{fig:sg_sym_c3v,fig:d6h} that the degree of bandmixing increases with increasing excitation levels, as expected. It is very low for the ground state, 99\% HH, and starts to be significant above level 6. 

We now proceed to identify the DPGPS, i.e. the dominating HSBF of a mixed state. Being HH-like, the DPGPS of all the first 15 Kramers conjugate pairs is $\ket{E_{3/2}^i}, i=1,2$ in \ctv, and the corresponding DPGPS in \dsh are the partners of $\ket{E_{3/2,g}}$ (c.f. \cref{app:dsh_hsbf}). 

The concepts of DPGPS and dominating UREF are important and will enable understanding of the quasi-degeneracies in the next section. 

\subsubsection{Analysis of UREFs}
\label{sec:urefs}
Let us first identify for every level (Kramers' pair) the dominant UREFs, using \cref{fig:d6h}. 
%We find the following dominances: level 1, 6 and 14 are $A_{1,g}$, levels 2 and 3 are $E_{1,u}$, levels 4 and 5 are $E_{2,g}$, level 7 is $B_{1,u}$, levels 8 and 9 are $E_{1,u}$, level 10 is $B_{2,u}$, level 11 is $E_{2,g}$, levels 12 and 15 are $E_{2,g}$, level 13 is $E_{2,u}$. 
We find the following dominances: level 1, 6 and 14 are $A_{1,g}$, levels 2, 3, 8, 9 and 13 are $E_{1,u}$, levels 4, 5, 11, 12 and 15  are $E_{2,g}$, level 7 is $B_{1,u}$ and  level 10 is $B_{2,u}$.
This is summarized in column 6 of \cref{tab:final_class}, column 4 (for \ctv) can be obtained by subduction. Note that we have already regrouped some levels together as pairs in this enumeration, the reason will become clear during the forthcoming analysis.

The concept of product states between the dominant UREF and DPGPS allows to explain most of the level sequence in \cref{fig:d6h}. Indeed, the double group irrep of each level (or the irrep pair in case of a level pair) can be faithfully generated by making the following products of irreps: for the isolated levels 1, 6 and 14 $A_{1,g} \times E_{3/2,g} = E_{3/2,g}$, for the level pairs 2, 3 and  8, 9 $E_{1,u} \times E_{3/2,g} = E_{5/2,u} + E_{1/2,u}$, for level pairs 4, 5 and 12, 15 $E_{2,g} \times E_{3/2,g} = E_{5/2,g} + E_{1/2,g}$, for level 7 $B_{1,u} \times E_{3/2,g} = E_{3/2,u}$, for level 10 $B_{2,u} \times E_{3/2,g} = E_{3/2,u}$, for level 11 $E_{2,g}\times E_{3/2,g} = E_{1/2,g}$ ($+ E_{5/2,g}$ but the latter is missing), and finally for level 13 $E_{1,u}\times E_{3/2,g} = E_{5/2,u}$ ($+ E_{1/2,u}$ but the latter is missing). The missing levels related to level 11 and to level 13 likely lie above the 15th energy level. The association of level 12 (instead of 11) with level 15 will be explained in the next section when looking at azimuthal and radial quantum numbers. It is still an open question whether the departure from \dsh symmetry of levels 11, 12 and 15 may be further minimized by including the missing partner of 11 in the symmetrization procedure (c.f. \cref{fig:unitary_c6v_dsh}).

The clustering in energy of the level pairs 2, 3 and  4, 5 and 8, 9, and 12, 15 (see \cref{fig:en_diff_c6v,tab:final_class}) can also be very clearly explained by dominant product states. It should be kept in mind that the quasi-degeneracies of 4,5 and 12,15 with levels 6 and 13,14 respectively are \textit{accidental} since these have different symmetries. Whilst the energy of the lower pairs are closely packed together due to negligible valence band mixing, higher pairs like 8, 9 and 12, 15 become more and more significantly split apart. The splitting is however already visible for the level pair 4, 5, manifested in \cref{fig:d6h} as non-vanishing weight in the central $E_{1/2}$ components in which there are distinct UREFs for the respective levels.  This is a clear demonstration that band-mixing is also a very relevant concept in the HSBF picture.

The isolated levels 1, 6 and 14 are all linked with the dominating $A_{1,g}$ UREF. Although it is customary that the fundamental level is strongly dominated by a fully invariant envelope function concentrated in the DPGPS component, the structure of levels 6 and 14 can again only be explained in the next section when looking at azimuthal and radial quantum numbers. Note that level 14 also display contributions from other HSBF components. 

Levels 7 and 10 are also interesting. They are very similar and in a way form a complementary pair because the relative strength of the weights of $B_{1u}$ and $B_{2u}$ is opposite between the two levels. In fact within one level none of the two is strongly dominant, but this is allowed by symmetry since the two products $B_{1,u} \times E_{3/2,g}$ and $B_{2,u} \times E_{3/2,g}$ give both of them global $E_{3/2,u}$ symmetry. In contrast with the partners of an $E_{1,u/g}$ or $E_{2,u/g}$ irrep (see \cref{fig:d6h}) which must be always balanced as predicted by the Wigner-Eckart theorem, the weights of the $B_{1u}$ and $B_{2u}$ UREFs must not necessarily be balanced in these levels. Levels 7 and 10 therefore illustrate the general fact that \textit{even if a state has a clear DPGPS} (clearly dominant HSBF) \textit{it may not necessarily have a single very dominant UREF}, as allowed by the analytical UREF decompositions given in \cref{app:uref_dsh}.  

Thus far we have only considered the DPGPS components of the spinors, however the higher excited levels also have a significant contribution from the $E_{1/2}$ HSBFs (LH). For simplicity we shall discuss this only in \ctv symmetry. First, note that all $E$-partner functions within LH components have balanced (equal) weight despite band mixing: this is imposed by symmetry (c.f. \cref{eq:vb_spinors}). By contrast, and in analogy to the discussion of the previous paragraph, level 13 has  $A_i, i=1,2$ UREF pairs in the LH components with an allowed weight imbalance. Second, we have seen analytically that time reversal symmetry imposes restrictions (c.f. \cref{eq:time_rev_req}) on the LH components of the $E_{1/2}$ states. As a clear example let us consider the first pair of $E_{1/2}$ states with significant LH weights, levels 8 and 9, which in addition contain all UREFs predicted by \cref{eq:vb_spinors}, as can be seen in \cref{fig:dg_sym_c3v_c3v_c6v}. We have been able to check numerically that indeed the relative phase of the $A_1$ and $A_2$ UREFs are out of phase with the $E$ UREFs by a factor $i$ in agreement with \cref{eq:time_rev_req}. Note also that level 12 has a rather large $A_1$ LH component, we do not think however that this state should be identified with a ground LH-like state. 

Finally, it is interesting to consider the parity of the dominant UREFs found with respect to the $\sigma_h$ operation. From the \dsh character table it is easy to see that all the single group irreps of the dominant UREFs appearing so far, namely $A_{1,g},A_{2,g},B_{1,u},B_{2,u},E_{2,g}$ and $E_{1,u}$ are even under $\sigma_h$ (nevertheless it should be noted that non-dominant UREFs may acquire some level of $z$ excitation due to bandmixing (c.f. \cref{fig:d6h})). The reason for this behavior of the dominant UREFs is rooted in the high lateral to axial aspect ratio of the QD under consideration which is 40:5 (c.f. \cref{fig:structure}). We therefore expect the envelope functions of the 15 lower levels considered here to have excitations mainly in the transverse plane (azimuthally and radially). This is investigated in the following subsection (\cref{sec:spatial}). 

To conclude we would like to stress that the dominant UREFs that we have identified in this subsection will play a key role not only in enabling the use of azimuthal and radial quantum numbers, but also in analyzing the strength of optical transitions.
%her%
\subsubsection{Approximate azimuthal and radial quantum numbers: the role of symmetry adapted functions}
\label{sec:spatial}
The goal of this section is to investigate the correspondence between a classical effective mass analysis for a disk-shaped quantum dot in terms of standard azimuthal and radial quantum numbers (as we have just seen there is no need for a vertical $n_z$ quantum number), and the sequence of irreps for the dominating UREFs identified in the previous section.  

The Schrödinger equation for a circular disk-shaped QD with infinite barriers is separable in azimuthal and radial coordinates, and leads to eigenfunctions which are proportional to products of an azimuthal exponential $\exp(i m_a \phi)$ with an $l_a$-th order Bessel function $J_{l_a}(k_{l_a,n_r}r)$ for the radial coordinate, where it is understood that $k_{l_a,n_r}R$ is the $n_r$-th zero of $J_{l_a}$ with $R$ being the dot radius. They are labeled by the azimuthal and radial quantum numbers, $l_a=0,1,\ldots,\infty$ and $n_r=1,\ldots,+\infty$. In cylindrical coordinates, $m_a$ is restricted to $m_a=\pm l_a$ (and axial excitations are labeled by $n_z$). The number of azimuthal nodes is thus $l_a$, and there is a twofold degeneracy for states with $l_a\neq0$.

\begin{figure}
	\centering
		\includegraphics[width=\columnwidth]{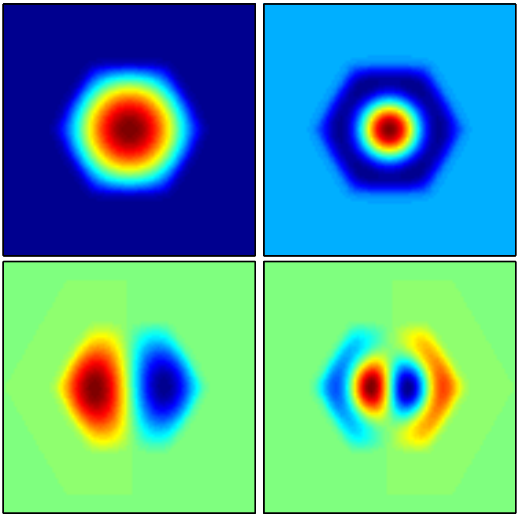}
		\caption{Cross sections of a selection of the contributions to the valence band spinor. The envelope functions shown in the top row are characterized as $\psi_{01}^{A_{1,g}}$ and $\psi_{02}^{A_{1,g}}$, in the order from left to right. The two functions in the lower row are $\psi_{11}^{E_{1,u},1}$  and $\psi_{12}^{E_{1,u},1}$.}  
	\label{fig:vb_cz_typical}
\end{figure}

Similar approximate quantum numbers can be expected in our nearly cylindrical 3D solutions, see \cref{fig:cb_3d} for the conduction band, and \cref{fig:vb_cz_typical} for the valence band spinor component cross sections. Therefore, in the following we shall specify the character of the computed dominating UREFs not only by their irreps, but also by additional subscripts $\psi^{\Gamma}_{n_a,n_r}$ where $n_a$ and $(n_r-1)$ specify the number of nodes in azimuthal and radial direction respectively (when they can be determined). The subscripts were identified by visual inspection for the first 15 levels, and are listed in the last column of \cref{tab:final_class}.  

We now consider the energy level sequences. First the sequence of levels 1, 6 and 14 with dominating UREF of $A_{1,g}$ symmetry: it is easy to see that they nicely form the sequence of $m_a=0$ levels with $n_r=1,2,3$, i.e. purely radial excitations. 

Another less trivial, but fundamental, sequence is the $n_r=1$, $l_a=1\ldots5$ sequence. We recognize the following five level pairs: 2 and 3, 4 and 5, 7 and 10, 11 and missing partner, 13 and missing partner. The question is now how to obtain further confirmation of this sequence. If one looks at the subduced representations from O(3), the so-called \textit{Symmetry Adapted Functions} (SAFs), listed in Table T35.5 and T35.6 of \ca, we can confirm that the link is correct, as the SAF quantum numbers $l,m,$ here corresponding to $l_a,m_a$ are in agreement with the appropriate single group labels. One finds that the $l=1,m=\pm 1$ basis subduces to $E_{1,u}$ in \dsh symmetry, $l=2,m=\pm 2$ subduces to $E_{2,g}$, $l=3,m=\pm 3$ subduces to $B_{1,u}+B_{2,u}$, $l=4,m=\pm 4$ subduces to $E_{2,g}$, and finally , $l=5,m=\pm 5$ subduces to $E_{1,u}$. Hence we have confirmed that this sequence of azimuthal excitations is in agreement with the sequence of irreps for the dominating UREFs, and additionally that levels 11 and 13 have missing partners which must be at higher energies. For level 11, this was also indicated by the structure of the intermixing blocks in the lower right corner of \cref{fig:d6h} as previously mentioned.

Similarly we can identify the last $n_r=2$, $l_a=1, 2$ sequence corresponding to level pairs 8, 9 and 12, 15. Again, the corresponding  $l,m$ quantum numbers subduce to the correct irreps $E_{1,u}$ and $E_{2,g}$ respectively. 

Even further interpretation is sometimes permitted. Let us consider again the pair of levels 7 and 10. Three nodal planes here correspond to either $B_{1,u}$ or $B_{2,u}$ symmetry. These are distinguished by a rotation by $\pi/6$ of the nodal planes. In our choice of coordinate system (\cref{fig:koord_dir}), the $\psi^{B_{1,u}}_{31}$ functions have nodal planes perpendicular to the facets of the hexagon, whereas the $\psi^{B_{2,u}}_{31}$, have nodal planes intersecting the corners of the hexagon. These two orientations give rise to different filling factors. The $\psi^{B_{1,u}}_{31}$ functions fill the area of the hexagon more effectively than the corresponding functions with $B_{2,u}$ labels. The curvature, and thus the energy is therefore higher for the $\psi^{B_{2,u}}_{31}$. This is also seen in the conduction band, as there is a splitting between the two corresponding levels (CB levels 7 and 8). For the valence band, the spinors do in fact contain mixtures of $\psi^{B_{1,u}}_{31}$ and $\psi^{B_{2,u}}_{31}$. Energy level 7 has $B_{1,u}$ as the largest contribution and a smaller contribution of $B_{2,u}$. The situation in energy level 10 is mirrored. The energy splitting between the levels is mostly due to the different curvature of $\psi^{B_{1,u}}_{31}$ and $\psi^{B_{2,u}}_{31}$. Note that the different curvature of states rotated by $\pi/4$ also apply for the $E_1/E_2$ states with azimuthal excitations. However for these states unequal weight in the UREF partners would violate the restrictions from Wigner-Eckart theorem.

As valence band mixing is weak for the lowest energy levels, the corresponding dominant HH envelopes will not hybridize with LH components. The lowest order UREFs observed in the LH components are $\psi^{B_{i,u}}_{31}, i=1..2$ envelope functions. This also further supports our previous assumption regarding the DPGPS as there is no sign of an LH-like sequence.

Note that the considerations of this section can also be carried out for the simpler conduction band, we leave this as an exercise for the reader. The situation in the valence band is much more complicated due to direction-dependent effective masses and valence band mixing. We expect that the approach also works when sequences of vertically excited states occur, and with $E_{1/2,g}$ DPGPS states (LH-like).

At this stage \Cref{tab:final_class} entails the full information for a complete classification of all states in compact form, both in true \ctv symmetry and in elevated \dsh symmetry. Note that the energy deviation due to symmetrization is the same in \dsh as in \ctv for the states labeled $E_{3/2,g}$ and $E_{3/2,u}$ (\dsh). Also note that in \dsh, the energy deviation is almost balanced within one pair. The last column of the table summarizes also the azimuthal and radial properties of the dominating UREF. 

In \cref{tab:final_class_cb} we present a similar summary for the conduction band states.

\begin{table*}
\caption{Final classification of valence band states.}
\label{tab:final_class}
\begin{tabular}{c@{}|@{\hspace{1pt}}c@{\hspace{1pt}}|@{\hspace{1pt}}c@{\hspace{1pt}}|@{\hspace{1pt}}c@{\hspace{1pt}}|@{\hspace{1pt}}c@{\hspace{1pt}}|@{\hspace{1pt}}c@{\hspace{1pt}}|@{\hspace{1pt}}c@{\hspace{1pt}}|@{\hspace{1pt}}c@{\hspace{1pt}}|@{\hspace{1pt}}c@{\hspace{2pt}}|@{\hspace{1pt}}c}
Level&\begin{tabular}{c}$E^\textrm{raw}$\\$[\textrm{meV}]$\footnotemark[1]\end{tabular} 
&\begin{tabular}{c}$\Delta E^{\textrm{\ctv}}$\\$[\muu eV]$\footnotemark[1]\end{tabular}
&\begin{tabular}{c}$\Delta E^{\textrm{\dsh}}$\\$[\muu eV]$\end{tabular}&$w_{LH}^{[111]}$\footnotemark[2]&
\begin{tabular}{c}$(\Gamma,\mu)^{\textrm{\ctv}}$\\\end{tabular}&
\begin{tabular}{c}UREF$^\textrm{\ctv}$\\dominant\end{tabular}&
\begin{tabular}{c}$(\Gamma,\mu)^{\textrm{\dsh}}$\\dominant\end{tabular}&
\begin{tabular}{c}UREF$^\textrm{\dsh}$\\dominant\end{tabular} &$\psi^{(\Gamma)^\textrm{\dsh}}_{(n_a,n_r)}$\\\hline
1&-14.6  &1.3&1.4&0.01&$E_{3/2}^j$&$A_1$&$E_{3/2,g}$&$A_{1,g}$&$\psi^{A_{1,g}}_{01}$\\  &&&&&&&&&\\
2,3&\begin{tabular}{c}-20.4,\\-20.9\end{tabular}&\begin{tabular}{c}0.4,\\0.5\end{tabular}&\begin{tabular}{c}118.2,\\-111.5\end{tabular}&\begin{tabular}{c}0.04,\\0.03\end{tabular}&$E_{1/2}$&$E$&\begin{tabular}{c}$E_{5/2,u}$,\\$E_{1/2,u}$\end{tabular}&$E_{1,u}$&$\psi ^{E_{1,u}}_{11}$\\&&&&&&&&&\\
4,  5&\begin{tabular}{c}26.4,\\  28.0\end{tabular}&\begin{tabular}{c}4.8,\\14.9\end{tabular}&\begin{tabular}{c}541.9,\\-487.5\end{tabular}&\begin{tabular}{c}0.12,\\0.08\end{tabular}&\begin{tabular}{c}$E_{1/2}$\end{tabular}&$E$&\begin{tabular}{c}$E_{5/2,g}$,\\  $E_{1/2,g}$\end{tabular}&\begin{tabular}{c}$E_{2,g}$\end{tabular}&$\psi ^{E_{2,g}}_{21}$\\  &&&&&&&&&\\
6&  -28.5  &-16.5&-15.9&0.15&$E_{3/2}^j$&$A_1$&$E_{3/2,g}$&$A_{1,g}$&$\psi ^{A_{1,g}}_{02}$\\  &&&&&&&&&\\
7,  10&\begin{tabular}{c}  -31.4,\\  -35.0  \end{tabular}&\begin{tabular}{c}6.3,\\ -9.5\end{tabular}&\begin{tabular}{c}8.9,\\ -12.1\end{tabular}&\begin{tabular}{c}0.25,\\ 0.21\end{tabular}&\begin{tabular}{c}$E_{3/2}^j$\end{tabular}&\begin{tabular}{c}$A_1$,\\ $A_2$ \end{tabular}&\begin{tabular}{c}$E_{3/2,u}$\end{tabular}&\begin{tabular}{c}$B_{1,u}$,\\ $B_{2,u}$ \end{tabular}&\begin{tabular}{c}$\psi ^{B_{1,u}}_{31}$,\\  $\psi ^{B_{2,u}}_{31}$\end{tabular}\\  &&&&&&&&&\\
8,  9&\begin{tabular}{c}  -33.9,\\   -34.6 \end{tabular} &\begin{tabular}{c}-2.8,\\ 8.5\end{tabular}&\begin{tabular}{c}319.8,\\ -296.3\end{tabular}&\begin{tabular}{c} 0.32\end{tabular}&$E_{1/2}$&$E$&\begin{tabular}{c}$E_{1/2,u}$,\\ \footnotemark[3] $E_{5/2,u}$\end{tabular}&$E_{1,u}$&$\psi ^{E_{1,u}}_{12}$\\  &&&&&&&&&\\
11&  -35.4 &1.2&285.0&0.37&$E_{1/2}$&$E$&\footnotemark[3]$E_{1/2,g}$&$E_{2,g}$&$\psi ^{E_{2,g}}_{41}$\\  &&&&&&&&&\\12,  15&\begin{tabular}{c}-37.8,\\-39.4  \end{tabular}&\begin{tabular}{c}7.5,\\-29.5\end{tabular}&\begin{tabular}{c}-57.7,\\-281.8\end{tabular}&\begin{tabular}{c}0.42,\\ 0.43\end{tabular}&\begin{tabular}{c}$E_{1/2}$\end{tabular}&$E$&\footnotemark[3]\begin{tabular}{c}$E_{1/2,g}$,\\  $E_{5/2,g}$\end{tabular}&\begin{tabular}{c}$E_{2,g}$\end{tabular}&\begin{tabular}{c}$\psi ^{E_{2,g}}_{22}$ \end{tabular}\\&&&&&&&&&\\
13&  -38.6& -3.5& -27.1&0.46&$E_{1/2}$&$E$&\footnotemark[3]$E_{5/2,u}$&$E_{1,u}$&$\psi ^{E_{1,u}}_{51}$\\  &&&&&&&&&\\14&  -39.0&16.3&15.7&0.44&$E_{3/2}^j$&$A_1$&$E_{3/2,g}$&$A_{1,g}$&$\psi ^{A_{1,g}}_{03}$
\end{tabular}
\footnotetext[1]{Averaged over Kramers doublets}
\footnotetext[2]{For the \ctv symmetrized basis}
\footnotetext[3]{Intermixed symmetries}
\end{table*}

\begin{table*}
\caption{Final classification of the conduction band states. The $(\Gamma,\mu)^{\textrm{\dsh}}$ labels are obtained taking the  direct product of the UREFs and the DPGPS representations (recall that spin-splitting is ignored). The DPGPS for all conduction band states is $E_{1/2,u}$(c.f.\cref{app:dsh_hsbf})}.
\label{tab:final_class_cb}
\begin{tabular}{c@{}|@{\hspace{1pt}}c@{\hspace{1pt}}|@{\hspace{1pt}}c@{\hspace{1pt}}|@{\hspace{1pt}}c@{\hspace{1pt}}|@{\hspace{1pt}}c@{\hspace{1pt}}|@{\hspace{1pt}}c@{\hspace{1pt}}|@{\hspace{1pt}}c}
Level&\begin{tabular}{c}$E^\textrm{raw}$\\$[\textrm{meV}]$\end{tabular}&$(\Gamma,\mu)^{\textrm{\ctv}}$&\begin{tabular}{c}UREF\\\ctv\end{tabular}&$(\Gamma,\mu)^{\textrm{\dsh}}$
& \begin{tabular}{c}UREF\\\dsh\end{tabular}&$\psi^{(\Gamma)^\textrm{\dsh}}_{(n_a,n_r)}$\\  &&&&&\\
1&1590&$E_{1/2}$&$A_1$&$E_{1/2,u}$&$A_{1,g}$  &$\psi^{A_{1,g}}_{01}$\\  &&&&&\\
2,3\footnotemark[2]&\begin{tabular}{c}1599,\\1599\end{tabular}&\begin{tabular}{c}$E_{3/2}^j$,\\$E_{1/2}$\end{tabular}&$E$&\begin{tabular}{c}$E_{3/2,g}$,\\$E_{1/2,g}$\end{tabular}&$E_{1,u}$&$\psi^{E_{1,u}}_{11}$\\  &&&&&\\
4,5\footnotemark[2]&\begin{tabular}{c}1612,\\1612\end{tabular}&\begin{tabular}{c}$E_{1/2}$,\\$E_{3/2}^j$\end{tabular}&$E$&\begin{tabular}{c}$E_{5/2,u}$,\\$E_{3/2,u}$\end{tabular}&$E_{2,g}$&$\psi^{E_{2,g}}_{21}$\\  &&&&&\\
6&1617  &$E_{1/2}$&$A_1$&$E_{1/2,u}$&$A_{1,g}$&$\psi^{A_{1,g}}_{02}$\\  &&&&&\\
7,  8&\begin{tabular}{c}1626,\\1630\end{tabular}&$E_{1/2}$&\begin{tabular}{c}$A_1$,\\$A_2$\end{tabular}&$E_{5/2,g}$&\begin{tabular}{c}$B_{1,u}$,\\$B_{2,u}$\end{tabular}& \begin{tabular}{c}$\psi ^{B_{1,u}}_{31}$,\\$\psi ^{B_{2,u}}_{31}$\end{tabular}\\  &&&&&\\
9, 10\footnotemark[2]&
\begin{tabular}{c}1636,\\1637\end{tabular}
&\begin{tabular}{c}$E_{1/2}$,\\$E_{3/2}^j$\end{tabular}&$E$&\begin{tabular}{c}$E_{1/2,g}$,\\$E_{3/2,g}$\end{tabular}&$E_{1,u}$&$\psi ^{E_{1,u}}_{12}$\\  &&&&&\\
\end{tabular}
\footnotetext[2]{Degenerate by symmetry. To enable better comparison with the valence band, we do however preserve individual level numbering for all 10 conduction band levels despite the degeneracy.}
\end{table*}
\section{Optical transitions}
\label{sec:opt_trans}

Sharp optical transitions in quantum dots are among their major attractive features for device applications. It is well-known that the oscillator strength corresponding to each interband $d$-polarized transition ($d=x,y,z$) at frequency $\hbar \omega = E^{\Gamma_\textrm{c}}_{n_\textrm{c}} - E^{\Gamma_\textrm{v}}_{n_\textrm{v}}$ is proportional to the summed squared $c-v$ interband matrix element $\bar{M}_{d,\,n_\textrm{c},n_\textrm{v}}(\Gamma_\textrm{c},\Gamma_\textrm{v})$, where the sum is over the degenerate contributing states, and can be written
\begin{align}
\lefteqn{\quad \bar{M}_{d,\,n_\textrm{c},n_\textrm{v}}(\Gamma_\textrm{c},\Gamma_\textrm{v}) =}  \label{eq:sME} \\ 
& & {\sum_{\Gamma_\textrm{c}'}}' {\sum_{\Gamma'_\textrm{v}}}' \sum_{\mu_\textrm{c},\mu_\textrm{v}} \left|\bra{\psi_\textrm{c}:n_\textrm{c},\Gamma'_\textrm{c},\mu_\textrm{c}} P_{d} \ket{\psi_\textrm{v}:n_\textrm{v},\Gamma'_\textrm{v},\mu_\textrm{v}}\right|^2 \nonumber.
\end{align}
The primed sum means a sum over eventually conjugated contributions, i.e. $\Gamma'$ summed over $\{\Gamma,\Gamma^*\}$ only if $\Gamma\neq\Gamma^*$. In \cref{eq:sME} $\psi_\textrm{c}$ and $\psi_\textrm{v}$ denote full conduction and valence band kets corresponding to energy levels  $E^{\Gamma_c}_{n_c}$ and $E^{\Gamma_v}_{n_v}$ respectively.

For the symmetry groups considered here, the set of momentum operators $P_{d}$, $d=x,y,z,$ can be divided into two distinct irreducible tensor operator (ITO) sets. In \ctv one has \cite{Altmann}\gbt{bevis??}:
\begin{equation} \label{eq_corC3v}
\left\{ P_x, P_y \right\} \longleftrightarrow \left\{ P^{E}_{\mu}, \mu=1,2 \right\} \quad , \quad P_z \longleftrightarrow P^{A_1}
\end{equation}
and in \dsh:
\begin{equation} \label{eq_corD6h}
\left\{ P_x, P_y \right\} \longleftrightarrow \left\{ P^{E_{1,u}}_{\mu}, \mu=1,2 \right\} \quad , \quad P_z \longleftrightarrow P^{A_{2,u}}
\end{equation}
The correspondences~(\ref{eq_corC3v}) and~(\ref{eq_corD6h}) allow the use of the generalized Wigner-Eckart theorem for evaluating the inner matrix element $\bra{\psi_\textrm{c}:n_\textrm{c},\Gamma_\textrm{c},\mu_\textrm{c}} P_{d} \ket{\psi_\textrm{v}:n_\textrm{v},\Gamma_\textrm{v},\mu_\textrm{v}}$. \mad{[removed primed here]}For our symmetry groups this leads to the prediction of optical oscillator strength isotropy in the $x-y$ plane~\cite{dupertuis_opt}, and to use group theoretical selection rules for optical transitions.

It should be stressed that the decomposition into UREFs linked with the HSBF basis allows to go beyond the simple use of the Wigner-Eckart theorem. Indeed, the UREFs lead straightforwardly to the prediction of "magic ratios" in polarization anisotropy~\cite{dupertuis_opt}, in the frame of our Luttinger model, and a conduction band \textit{with no spin-splitting}. To this end, full conduction band spinors must be reconstructed by adding spin as in Ref.~\citenum{dupertuis_opt}. In the case of \ctv symmetry, four types of Kramers degenerate pairs then appear : $E_{1/2}(A_1)$ and $E_{1/2}(A_2)$ for conduction band states with $A_1$ and $A_2$ envelopes respectively, and $E_{1/2}(E)$ and $E_{3/2}(E)$ for conduction band states with $E$ envelopes. The latter two are always degenerate.
Then, the polarization anisotropy, defined as $A_{ij}=(\bar{M}_i-\bar{M}_j)/(\bar{M}_i+\bar{M}_j), i,j \in \{x,y,z\}$ can be calculated using UREFs, and reveals the magic anisotropy ratios $A_{zx}$ given in \cref{eq:sel_rules_ar}. The ratios $\pm1$ in \cref{eq:sel_rules_ar} stem from standard \ctv selection rules. The Wigner-Eckhart theorem also predicts $A_{xy}=1$ and $A_{zy} = A_{zx}$ for all irreps. 

\begin{table}[htbp]
\caption{Optical transition anisotropy ratios, $A_{zx}$, for QD's with \ctv symmetry}\label{eq:sel_rules_ar}
	\centering
		\begin{tabular}{c|cc}
\backslashbox{$\Gamma_\textrm{c}$}{$\Gamma_\textrm{v}$}&$^iE_{3/2}$&$E_{1/2}$\\\hline
$^iE_{3/2}(E)$&1&-1\\
$E_{1/2}(E)$&-1&$f(\psi^{E}_\textrm{c},\Phi_\textrm{v}^{E},\phi_\textrm{v}^E)$\footnotemark[1]\\
$E_{1/2}(A_i)$&-1&$\frac{3}{5}$\\
&&
		 \end{tabular}
\footnotetext[1]{$f(\psi^{E}_\textrm{c},\Phi_\textrm{v}^{E},\phi_\textrm{v}^E)=\frac{\frac{2}{3}\norm{\psi^{E}_\textrm{c}}{\Phi_\textrm{v}^{E}}-\left|\Re\braket{\psi^E_\textrm{c}}{\phi_\textrm{v}^E}\right|^2}{\frac{2}{3}\norm{\psi^{E}_\textrm{c}}{\Phi_\textrm{v}^{E}}+\left|\Re\braket{\psi^E_\textrm{c}}{\phi_\textrm{v}^E}\right|^2}$}
\end{table}

In the present work we shall now show that when the identification of the dominant UREF can be performed, one obtains further clues to the oscillator strength spectrum. Moreover, when azimuthal and radial quantum numbers can also be identified, further constraints occur. In both cases the presence of main lines can be explained, and fine-structure can be interpreted.
%\subsection{Numerical calculation of \texorpdfstring{\ctv}{ctv} optical oscillator strengths}
\subsection{Numerical calculation of \ctv optical oscillator strengths}
\begin{figure*}
	\centering
			\includegraphics[width=\textwidth]{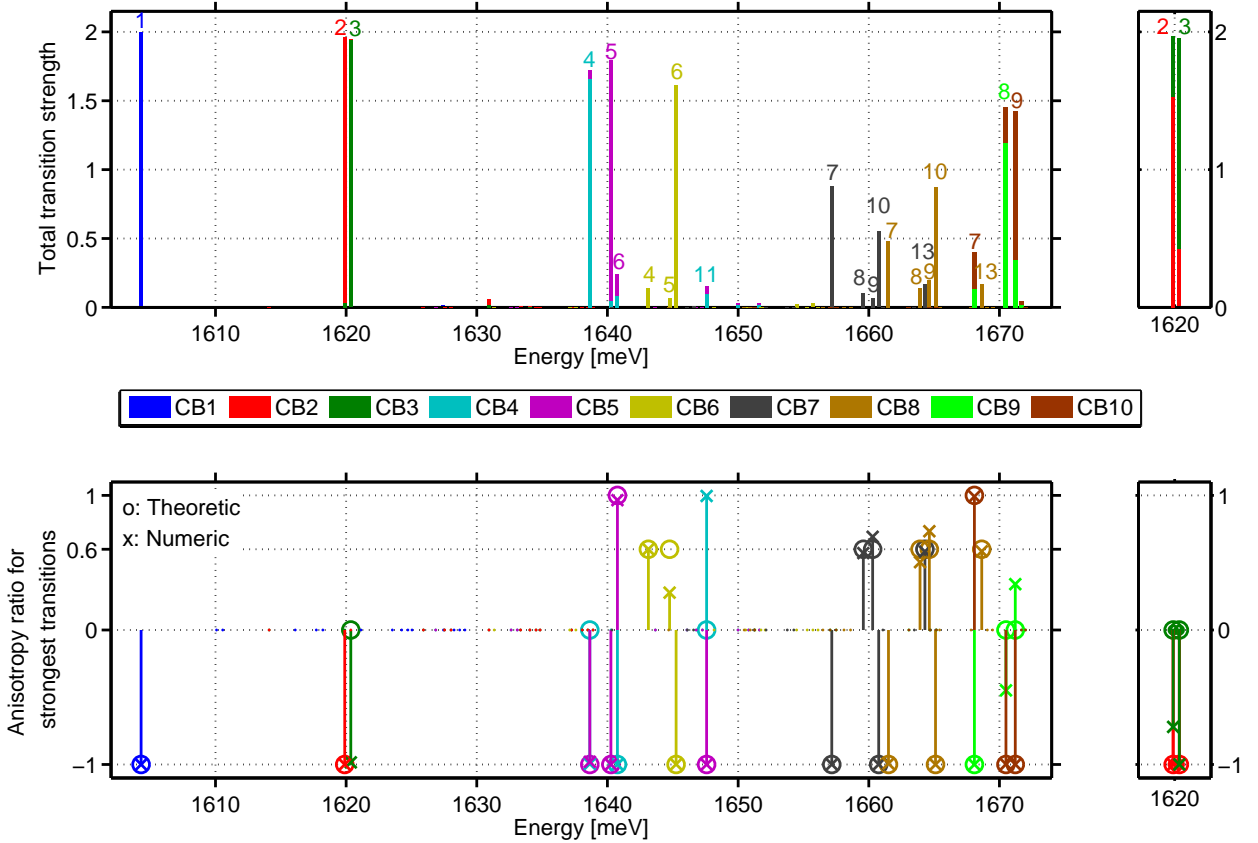}
		\caption{The top subplot show the combined optical oscillator strength, $\bar{M}_x+\bar{M}_y+\bar{M}_z$, as a function of energy. The colors label the contributions due to the individual conduction band levels, and the valence band state number is indicated by numbers. The lower subplot show the anisotropy ratios, $A_{zx}$, for the significant transitions (limit set to $\bar{M}_x+\bar{M}_y+\bar{M}_z$=0.05). The theoretical anisotropy ratio predictions are also included. Some of the anisotropy ratios cannot be predetermined, for these transition the theoretical value has been set to the dummy value 0 to mark  them. The inset at the right show the $CBi-VBj, i,j=2,3$ in the \ctv basis.}
	\label{fig:opt_trans}
\end{figure*}

The calculated optical oscillator strengths of the nanowire QD are displayed as a function of the transition frequency in \cref{fig:opt_trans}. We used \dsh symmetrized valence band states since it allows to interpret the optical spectra in most details. In the side figure the details of two transitions calculated in the \ctv-symmetrized basis are also displayed, other differences were minor. By contrast, for the conduction band we used the raw calculated states since these states were already sufficiently symmetrized, rendering unnecessary a further \dsh symmetrization. Nevertheless we restored in \cref{fig:opt_trans} an exact degeneracy for $E$ conduction band states, for the purpose of clarity of the fine structure in the figure. The numerical splitting of the computed energies was anyway below the estimated convergence (cf. \cref{fig:cb_conv}).

The upper subplot in \cref{fig:opt_trans} shows the total oscillator strengths for the main transitions (summed over in all directions), whilst the lower subplot shows the corresponding $A_{zx}$ anisotropy ratios. This data suffices for our purpose, but the reader may recover the separate results in each direction using the analytical results $A_{xy}=1$ and $A_{zy} = A_{zx}$,  whose  validity was also confirmed numerically. It must be understood that other effects also occur in a real experiment, besides the neglected Coulomb contributions. They may change to some extent the predicted spectrum, in particular in nanowires there are important effects due to the high index contrast between the nanowire and the air \cite{moses_pl_aniso} (dielectric mismatch effect). Nevertheless, the oscillator strength spectra given by single particle calculations as in \cref{fig:opt_trans} are often the main characteristic of the intrinsic optical response of QDs.

Let us now study more in details the oscillator strengths of each optical transition, which is color coded in \cref{fig:opt_trans}. They are numbered so that the properties of the conduction band level $i$ and valence band level $j$ corresponding to transition $CBi-VBj$ can be directly read off using the previously obtained \cref{tab:final_class,tab:final_class_cb}, and Eqs.~(\ref{eq_corC3v}) or~(\ref{eq_corD6h}). 

\subsection{Dominant and missing optical transitions}
A first striking feature of the upper part of \cref{fig:opt_trans} is the dominant diagonal character of the optical transitions, i.e. $CBi-VBj$ is most intense for $i=j$. Note that within degenerate CB levels the numbering has been explicitly chosen to respect $CBi-VBj$ diagonality to ease the analysis. This diagonal character is nearly perfect for the ground state transition, as well as for the first pair $i=2,3$, but progressively weakens as one climbs the excitation ladder.

We also clearly see in \cref{fig:opt_trans} that the highest pairs give rise to a richer structure with side peaks due to the valence-band complexity, but still the $(CB7,CB8)$ pair is dominantly diagonally coupled to the $(VB7,VB10)$ pair, and $(CB9,CB10)$ to $(VB8,VB9)$ pair (see \cref{tab:final_class}). Clearly, cross-coupling like CB7-VB10, e.t.c. also appear as sub-structures within these dominant pairs.

Looking back to the character of the corresponding states in \cref{tab:final_class,tab:final_class_cb}, we find that symmetry elevation to \dsh is an important ingredient to explain missing non-diagonal transitions in \cref{fig:opt_trans}. However, not all of them are explained. Take for example the set of transitions from the ground conduction band level $CB1-VBj$. Only $CB1-VB1$ is observed. In principle, using double group selection rules in \ctv, all values of $j$ are permitted from $CB1$ in $x-y$-polarization, but in \dsh only $j=1,5,6,11,12,14$ are permitted. In $z$-polarization $j=2-5,8,9,11-13,15$ are permitted in \ctv whilst only $j=5,11,12$ would be in \dsh. Thus symmetry elevation accounts for part of the missing transitions, but obviously it is not enough to explain all of them.

Actually one should consider in addition \textit{both} the symmetry of the dominant UREF and the presence of the approximate azimuthal and radial quantum numbers to explain all missing transitions for the lowest energy states. This is particularly clear with the set $CB1-VBj$. First, on the basis of the symmetry of the dominant UREF $A_{1,g}$ of $CB1$ and the contributing UREFs of $VBj$, all $z$-polarized lines are forbidden, and only $j=1,6, 14$ are allowed in $x-y$ polarization. Second, to explain the remaining missing lines from $CB1$, corresponding to $j=6,14$, it suffices to invoke the \textit{approximate} selection rule based on the radial quantum number of the dominant UREF, which is $n_r = 1$ for $CB1$, and $n_r = 2, 3$ for $VB6$ and $VB14$ respectively (see \cref{tab:final_class,tab:final_class_cb}).

Let us examine more closely the optical coupling between diagonally coupled pairs,  considering the UREFs. In $(CB2,CB3)-(VB2,VB3)$, symmetry elevation to \dsh is not sufficient to predict only $CB2-VB2$ and $CB3-VB3$ transitions as observed in \cref{fig:opt_trans}. In this respect, it is interesting to see that if one assumes \ctv symmetry, both the crossed transitions between the pairs ($CB2-VB3$ and $CB3-VB2$) should appear, and indeed this is what we observe in the calculated spectrum in the \ctv-symmetrized basis (see side of \cref{fig:opt_trans}). The observation of $CB3-VB2$ would witness the \dsh-symmetry breaking present in the [111] Luttinger Hamiltonian, and might well be resolvable experimentally, whilst CB2-VB3 would be compatible with both symmetries.

In the higher diagonally coupled pairs $(CB4, CB5)-(VB4, VB5)$ and $(CB9, CB10)-(VB8, VB9)$ the same effects are observed and the same considerations apply. For these two sets of transitions we also see in \cref{fig:opt_trans} the appearance of additional non-diagonal transitions with other levels, a trend which naturally increases with transition energy. The last diagonally coupled pairs $(CB7,CB8)-(VB7,VB10)$ are particularly interesting because the two electronic levels $CB7$ and $CB8$ are non-degenerate, hence in \cref{fig:opt_trans} one observes four dominant peaks for these transitions.

\subsection{Optical anisotropy of dominant transitions}

The optical anisotropies are also of fundamental interest. By contrast to the selection rules discussed in the previous section, nearly all optical anisotropies can be explained in \ctv symmetry. 

\Cref{eq:sel_rules_ar}, with the help of the classification of \cref{tab:final_class,tab:final_class_cb}, predicts the anisotropy ratio to be $-1$ for transitions between isolated levels $CBi-VBi,\, i=1\textrm{ or } 6$, in agreement with the lower part of \cref{fig:opt_trans}. 

The transitions between the coupled pairs $(CB7,CB8)$ and $(VB7,VB10)$ are also predicted to have anisotropy ratio $-1$, as is well verified in \cref{fig:opt_trans}. However the situation for  the other diagonally coupled pairs is more subtle. 

In $(CB2,CB3)-(VB2,VB3)$, the optical anisotropy is predicted to be undetermined for the two transitions stemming from $CB3$, and $-1$ for the two stemming from $CB2$. This is satisfied in the \ctv-symmetrized basis (see side of \cref{fig:opt_trans}), but the anisotropy of both, particularly $CB3-VB3$, is still close to $-1$. The explanation is very simple if one recalls that the LH weight in $VB2$ and $VB3$ is less than 4\% (from \cref{eq:sel_rules_ar}, the undetermined anisotropy ratios will be close to	-1 if the overlap with $\Phi^E_v$ vanishes). The departure from $-1$ is much bigger for $CB3-VB2$, which can be understood by recalling that this transition should be forbidden in the $x-y$ polarization in \dsh \mad{[this is not clear to me, as also z polarization forbidden in\dsh?]}, hence it is more sensitive to LH admixture. We should point out that the disappearance of the transition $CB2-VB3$ in the \dsh-symmetrized basis, evidenced in \cref{fig:opt_trans}, cannot be understood on the basis of \dsh symmetry alone. We suspect that a study generalizing the magic ratios to \dsh symmetry, linked with the neglect of conduction band spin-splitting, might explain this observation, but the analytical verification would be quite overwhelming. 

The optical anisotropy of $(CB4,CB5)-(VB4,VB5)$ and $(CB9,CB10)-(VB8,VB9)$ is entirely similar to $(CB2,CB3)-(VB2,VB3)$ but with the role of the two conduction and valence band levels reversed. We also see that the undetermined anisotropy ratio in transitions from $CB9$ to $(VB8,VB9)$ depart more significantly from any specific values, as is expected due to increased band mixing.

\subsection{Fine structure due to valence-band mixing}
Besides the series of dominant peaks there is a fine structure produced by valence band mixing  which is particularly interesting to investigate. \cref{fig:opt_trans} shows that such band mixing really starts to be significant from the set of optical transitions $CBi-VBj$ with $i,j\geq 4$.

From the pair $(CB4,CB5)$, weak additional transitions to $VB6$ and $VB11$ are visible in \cref{fig:opt_trans}. From the Wigner-Eckart theorem, the transitions to $VB6$ are both allowed in elevated symmetry \dsh whilst towards $VB11$ only $CB5-VB11$ is allowed. We conclude that $CB4-VB11$ is a manifestation of the true \ctv dot symmetry. These weak transitions are slightly more intense in $z$-direction. The anisotropy ratios for the transitions $CB4-VB6$ and $CB5-VB11$ are -1, and for $CB5-VB6$ it is +1, in agreement with \ctv symmetry (\cref{eq:sel_rules_ar}). For the remaining $CB4-VB11$ transition the undetermined anisotropy ratio proves to be numerically close to $+1$. Clearly, looking at the corresponding $E_{2,g}$ UREFs, in the spinors \cref{eq:psi_1_E_32u,eq:psi_2_E_32u} we may confirm that both +1 transitions \textit{are purely due to valence band mixing} since only the $E_{1/2,g}$ HSBF (LH-components) can couple to $CB5$.

Besides the next dominant transition $CB6-VB6$, we can see two weak transitions  $CB6-VB4$ and $CB6-VB5$ in \cref{fig:opt_trans}. Again, both transitions are due to valence band mixing, and are not predicted in elevated symmetry \dsh; they manifest the true \ctv symmetry of the QD. The novelty here is that both transitions should have the "magic" anisotropy ratio $3/5$ predicted by \cref{eq:sel_rules_ar}. This is very well satisfied by $CB6-VB4$, but $CB6-VB5$ displays a numerical anisotropy ratio close to $0.3$ which will represent the largest deviation from the predicted value in our numerical data. Since its oscillator strength is already rather weak we can safely attribute this deviation to imperfect \ctv symmetrization. 

Let us now investigate the weak transitions from the $(CB7,CB8)$ pair to levels $VBj,\, j=8,9,13$. Again these six transitions due to valence band mixing manifest the true \ctv symmetry of the QD, and display quite accurately the ``magic'' anisotropy ratio $3/5$  for $E_{1/2}(A_1)-E_{1/2}$ transitions.

The last weak transitions are related to the highest excited pair $(CB9,CB10)$ towards $VB7$ and display polarization anisotropy ratios of $(+1,-1)$ respectively. The symmetry considerations are the same as for $(CB4,CB5)$ to $VB6$ (but correspond to a higher radial excitation), so they are both allowed in \dsh symmetry, and $CB9-VB7$ is a LH transition purely due to valence band mixing.

\subsection{Summary of optical transition spectrum}
Optical transitions with polarization perpendicular to the nanowire axis are clearly more dominant for the lowest energy states. This is expected due to the oblate (quantum-well-like) aspect ratio of the QD (\cref{fig:structure}), the other optical transitions seen with optical activity polarized along the nanowire axis are linked with valence band mixing and overlap between the conduction band and the LH components. Accordingly, we see an increasing number of transitions with this polarization for higher levels, as the valence band states become increasingly LH-like (see \cref{tab:final_class}). A computation with more energy levels, or a narrower QD with stronger lateral confinement, would have allowed to reach the ground LH-like level with strong dominance of $z-$polarized optical transitions. Our approach would then evidence the second kind of DPGPS. If the QD aspect ratio would be reversed to prolate a reversal of the roles of HH and LH is expected. 

Excitonic effects, which we have neglected in the present work, are not expected to change very much this global picture of polarization anisotropy. It would introduce electron-hole exchange effects which would split all the dominant transitions into a further observable fine structure (which can be seen as quadruplets due to spin degeneracy). The symmetries of the fine-structure excitons can be easily obtained by the product of irreps \cite{Dupertuis_letter}. Sometimes doublets may remain, other times all lines would split in \dsh. 

In a real situation, note that the dielectric mismatch between the high index nanowire and the surroundings may filter emission along the nanowire axis, altering the intrinsic optical response of the nanowire QD discussed here. 

\section{Conclusion}
\label{sec:conc}
We have presented a systematic procedure based on class operators for symmetry analysis of the electronic states of a QD. The procedure, called PTCO, is based on postprocessing and alleviates the need for a code specialized with respect to a given symmetry (which would however feature significant gains in memory/time \cite{dupertuis_gallinet}). PTCO has been demonstrated using the \kp method for the conduction and valence band of hexagonal GaAs QD grown within \algaas nanowires. The high \dsh symmetry of the QD heterostructure is partly broken by the \ctv crystal symmetry carried by the Luttinger Hamiltonian, and by the unsymmetric computing grid. Using the PTCO on the computed results, we have been able to sort all these symmetry breaking effects, and quantify them. We have demonstrated that the numerical grid effect was small and could be compensated by PTCO, whilst the deviations from the approximate elevated \dsh symmetry towards the true \ctv symmetry, albeit small, could be measurable.

The PTCO is simple to program, intrinsically systematic and automatized, and is carried out in a single step for a given symmetry, delivering at the same time symmetrized states with corrected energies and a classification for all quantum states. It is flexible, applies independently of the method used to compute the electronic structure, and can be tuned with very little efforts to analyze a higher symmetry in a second run. This enabled us to investigate the proximity of each state to an approximate elevated symmetry.

In a second step, all the quantum states were  analyzed using projection operators to give quantitative weights for every symmetry group, for every ultimately reduced envelope function present in every spinorial component. We could then verify all analytical predictions made concerning the UREFs in the elevated \dsh symmetry (\cref{app:uref_dsh}). This approach allowed to identify dominant DPGPS and dominant UREFs for all states considered.

The analysis of the dominant UREFs opened the possibility to attribute additional azimuthal and radial quantum numbers to every state. The natural sequence allowed to explain the order of irreps in the computed results, by subduction from $O(3)$ to \dsh. 

A final summary of the classification is given for all valence band states in \cref{tab:final_class}, and for all conduction band states in \cref{tab:final_class_cb}. The classification provides insights into the origin of degeneracies and quasi degeneracies, and allows to predict all selection rules and most of the polarization properties. The information can also be used to construct approximate product states, relevant for the interpretation of excitonic and multi-excitonic fine structure \cite{Dupertuis_letter}. 

Finally we were able to interpret completely all the details of the computed optical spectrum with the help of the classification of the states. We unveiled a large number of missing transitions, which were shown to stem not only from approximate elevated symmetry but also from approximate azimuthal and radial quantum numbers. As valence band mixing was included in the model, we could identify all the corresponding fine structure, responsible for some spread in the optical transition spectrum, and to verify analytical predictions concerning the appearance of "magic ratios" in polarization anisotropy. 

High symmetry GaAs nanowire QDs as investigated here are particularly interesting. First, since there are a large number of forbidden transitions in their optical spectrum, the multi-excitonic spectrum is also simplified and can be better interpreted. Second, the doubly degenerate exciton states due to symmetry are well suited for the generation of entangled photons. \gbt{hvilken?}

\begin{acknowledgments}
This work was supported by the "NANOMAT" program (grant No 182091) of the Research Council of Norway. 
\end{acknowledgments}
\appendix
%\section{Choice of standard irreducible matrix representations for \texorpdfstring{\ctv}{c3v} and \texorpdfstring{\dsh}{d6h}}
\section{Choice of standard irreducible matrix representations for \ctv and \dsh}
\label{app:irreps_ctv_dsh}
For one dimensional irreps $\Gamma$, the set of matrices $\{D^\Gamma(g),g\in\mathcal{G}\}$ are simply the characters of the respective irreps i.e. $D^\Gamma(g),\chi^\Gamma{g},\forall g$, where $\chi^\Gamma(g)$ is listed in \citenum{Altmann}. When $\Gamma$ is of higher dimension, the matrices  $\{D^\Gamma(g)$ are only unique up to a similarity transform $U(\Gamma)$, i.e.
\begin{equation}
D^\Gamma(g)=U(\Gamma)^{-1}\left[D_\textrm{Alt}^\Gamma(g)\right]^{H}U(\Gamma)\label{eq:sim_rel_irrep}
\end{equation}
where $D_\textrm{Alt}^\Gamma(g)$ is a matrix irrep listed in \citenum{Altmann},and $U(\Gamma)$ is restricted to be unitary. For our purpose the $\left[D_\textrm{Alt}^\Gamma(g)\right]^{H}$ are not the most numerically convenient since they diagonalize rotations instead of a mirror operation w.r.t. a symmetry plane, e.g. $\sigma_{v1}$ (\cref{fig:koord_dir}). We thus choose the following to define our $D^\Gamma(g)$ matrices:\\
\ctv, irrep $E$:
\begin{eqnarray}
U_\textrm{\ctv}(E)=
\left[\begin{array}{cc}
   \frac{i}{\sqrt{2}}&\frac{1}{\sqrt{2}}          \\-\frac{i}{\sqrt{2}}&\frac{1}{\sqrt{2}}\end{array}\right]
\label{eq:U_ctv_E}
\end{eqnarray}\\
\ctv, irrep $E_{1/2}$ (with case A in \ca):
\begin{eqnarray}
U_\textrm{\ctv}(E_{1/2})=
\left[\begin{array}{cc}
   \frac{i}{\sqrt{2}}&\frac{-1}{\sqrt{2}}          \\\frac{1}{\sqrt{2}}&\frac{-i}{\sqrt{2}}\end{array}\right]
\end{eqnarray}
\dsh, irreps $E_i$: 
\begin{equation}
U_\textrm{\dsh}(E_i)=U_\textrm{\ctv}(E), \,\, i=1..2
\end{equation}
\dsh, irreps $E_{j,k}$: 
\begin{align}
&U_\textrm{\dsh}(E_{j,k})=U_\textrm{\ctv}(E_{1/2}), \,\, j=\frac{1}{2},\frac{5}{2},k=u,g
\label{eq:U_dsh_12}\\
&U_\textrm{\dsh}(E_{3/2,k})=
\left[\begin{array}{cc}
   \frac{i}{\sqrt{2}}&\frac{1}{\sqrt{2}}          \\\frac{-1}{\sqrt{2}}&\frac{-i}{\sqrt{2}}\end{array}\right],\,\, k=u,g
\label{eq:U_dsh_32}
\end{align}
 Note first that our matrix representation obey transposed multiplication tables w.r.t. \ca (c.f. Hermitian conjugation in \cref{eq:sim_rel_irrep}, this is required by our use of the passive rather than the active point of view). Second, our special choice \cref{eq:U_dsh_32} for $E_{3/2}$ in \dsh stems from separate requirements concerning the "optimal" choice of HSBF. We ensured in particular that the resulting $E_{3/2,g}$ valence band HSBF was simultaneously the same for \dsh and \ctv symmetry.
 
%\section{HSBF for the conduction and valence band spinord in elevated \texorpdfstring{\dsh}{dsh} symmetry}
\section{HSBF for the conduction and valence band spinord in elevated \dsh symmetry}
\label{app:dsh_hsbf}
Symmetrized bases \cite{Altmann}, also called \textit{Symmetry Adapted Functions} (SAF), follow from subduction from $O(3)$ to \dsh and allow to find easily the proper HSBF basis corresponding to the top (bottom) of the valence (conduction) band, respectively, of diamond or Zinc Blende. Unfortunately the SAF in \ca do not have regular properties under time-reversal, so some care is required.

Let us first consider the bottom of the conduction band, and use linear combinations of zone center Bloch function denoted $\ket{\frac{1}{2},m}^\bullet$, $m=\pm\frac{1}{2}$, corresponding to a "quantized axis" with  $z$ along [111] and $x$ along $[11\bar{2}]$ \cref{fig:koord_dir}. They are nearly odd under spatial inversion in GaAs (hence the $\ket{\dots}^\bullet$ symbol as in \ca). Subduction from $O(3)$ tells us that the irrep $E_{1/2,u}$  of \dsh must be associated with this subspace. To construct the HSBF we can either use the SAF of \ca which are suitable in this case, and the change of basis corresponding to \cref{eq:U_dsh_12}, or equivalently decide to diagonalize the set of Wigner matrices corresponding to the symmetry operations (parametrization and factor system of \ca), and find their reduction to the set $D^{E_{1/2,u}}(g)$ given in \cref{app:irreps_ctv_dsh}. The resulting conduction band HSBFs are:
\begin{align}
\left.\begin{array}{l}
\ket{E_{1/2,u},1}=-\frac{i}{\sqrt{2}}\ket{\frac{1}{2},\frac{1}{2}}^\bullet-\frac{1}{\sqrt{2}}\ket{\frac{1}{2},-\frac{1}{2}}^\bullet\\
\ket{E_{1/2,u},2}=\frac{1}{\sqrt{2}}\ket{\frac{1}{2},\frac{1}{2}}^\bullet+\frac{i}{\sqrt{2}}\ket{\frac{1}{2},-\frac{1}{2}}^\bullet
\end{array}\right\}
\label{eq:cb_hsfb_dsh}
\end{align}

For the top of the valence band we use the set of similar Bloch-functions labeled $\ket{\frac{3}{2},m}, \,m=-\frac{3}{2}...\frac{3}{2}$, \gbt{I have one question regarding the HSBF appendix: Is it supposed to be  -3/2...3/2 or oppositely?}nearly \textit{even} under inversion. Then the $O(3)$ subduction tables to \dsh indicate reduction to the $E_{3/2,g}+E_{1/2,g}$ irreps. This time one cannot use all the SAFs of \ca if one wants to keep an invariant form of the time reversal operator in the HSBF, therefore we follow the route of reducing the Wigner matrices corresponding  to symmetry operations. Using a suitable, but freely chosen, set of phase factors compatible with the irreps of  \cref{app:irreps_ctv_dsh}, one finds the following \dsh valence band HSBFs:
\begin{equation}
\left.\begin{array}{c}
\ket{E_{3/2,g},1}=\frac{i}{\sqrt{2}}\ket{\frac{3}{2},\frac{3}{2}}		-\frac{1}{\sqrt{2}}\ket{\frac{3}{2},-\frac{3}{2}}\\
\ket{E_{3/2,g},2}=\frac{1}{\sqrt{2}}\ket{\frac{3}{2},\frac{3}{2}}-\frac{i}{\sqrt{2}}\ket{\frac{3}{2},-\frac{3}{2}}
\end{array}\right\}
\label{eq:vb_hsfb_dsh_32}
\end{equation}
and
\begin{equation}
\left.\begin{array}{c}
\ket{E_{1/2,g},1}=\frac{1}{\sqrt{2}}\ket{\frac{3}{2},\frac{1}{2}}		+\frac{i}{\sqrt{2}}\ket{\frac{3}{2},-\frac{1}{2}}\\
\ket{E_{1/2,g},2}=\frac{i}{\sqrt{2}}\ket{\frac{3}{2},\frac{1}{2}}+\frac{1}{\sqrt{2}}\ket{\frac{3}{2},-\frac{1}{2}}
\end{array}\right\}
\label{eq:vb_hsfb_dsh_12}
\end{equation}
This choice satisfies three desirable constraints: 1) the matrix form of the time reversal operator \notPart{\cref{eq:time_rev}} is invariant \notPart{w.r.t. the $\{\ket{\frac{3}{2},m}\}$ basis} when the valence band basis is ordered as 
\begin{equation}
\{\ket{E_{3/2,g},1},\ket{E_{1/2,g},1},\ket{E_{1/2,g},2},\ket{E_{3/2,g},2}\}
\label{eq:vb_dsh_hsbf_basis}
\end{equation}
hence it preserves the canonical $p,q,r,s$ form (\cref{eq:vb_ham}) of the Luttinger Hamiltonian in this basis, 2) it is also simultaneously a HSBF for \ctv, 3) their transformation laws are given by our set of standard representations listed in \cref{app:irreps_ctv_dsh}.

For \ctv the choice \cref{eq:vb_hsfb_dsh_12,eq:vb_hsfb_dsh_32} is in agreement with Ref.~\citenum{dupertuis_opt} with the correspondence 
\begin{equation}
\ket{E_{3/2,g},i}\to\ket{^iE_{3/2}},\, i=1..2
\label{eq:vb_hsbf_dsh_opt}
\end{equation}
whilst it differs from Ref.~\citenum{dupertuis_msr} by

\begin{equation}
\left.\begin{array}{l}
\ket{E_{3/2,g},i}\to(-1)^{i}\ket{^{3-i}E_{3/2}}\\
\ket{E_{1/2,g},i}\to(-1)^{i}\ket{E_{1/2},3-i}
\end{array}\right\},\, i=1..2.
\label{eq:vb_hsbf_dsh_msr}
\end{equation}

Although the matrix representation for \ctv in \cref{app:irreps_ctv_dsh}  are the same as in Ref.~\citenum{dupertuis_msr}, we had chosen an opposite projective factor system for the improper operations in Ref.~\citenum{dupertuis_msr}. The present projective factor system is now the same as in \ca.
%\section{UREF decompositions of the valence band spinors in the elevated \texorpdfstring{\dsh}{dsh} symmetry}
\section{UREF decompositions of the valence band spinors in the elevated \dsh symmetry}
\label{app:uref_dsh}
Using the HSBF derived in \cref{app:dsh_hsbf}, \cref{eq:vb_hsfb_dsh_12,eq:vb_hsfb_dsh_32,eq:vb_dsh_hsbf_basis}, it is possible to decompose each spinor into UREFs with the help of Eq.~(48) of Ref.~\citenum{dupertuis_msr} and the Clebsch-Gordon coefficients linked with the chosen standard matrix representation of \cref{app:irreps_ctv_dsh}, and involving a minimum of arbitrary phase factors. In case of \ctv symmetry one finds that the UREF decomposition of the main text Eqs.~\eqref{eq:vb_spinors}, which remain identical to Refs.~\citenum{dupertuis_msr,dupertuis_opt}. In case of \dsh symmetry, and in the HSBF basis order given by \cref{eq:vb_dsh_hsbf_basis} below, one obtains the ungerade valence band spinors given by \cref{eq:vb_spinors_dsh}. The corresponding gerade spinors have similar expressions where $u\to g$. To lighten the notation we have omitted the main spinor index and the HSBF index in the UREFs, but they can be retrieved easily from the main spinor and the position of the UREF (using \cref{eq:vb_dsh_hsbf_basis}). For clarity the $\phi/\Phi$ UREFs are not the same functions in 
Eqs.~\eqref{eq:psi_1_E_32u}-\eqref{eq:psi_2_E_32u} w.r.t. Eqs.~\eqref{eq:psi_1_E_12u}-\eqref{eq:psi_2_E_12u}, and we have used the capital $\Phi$ to distinguish the $E_1$ and $E_2$ UREFs in the $E_{1/2,g}$ HSBF components from the $E_{3/2,g}$ HSBF components.

Subduction rules to \ctv (according to the case A in \ca) will clearly reduce the spinors \cref{eq:vb_spinors_dsh} to \cref{eq:vb_spinors}. Since $E_1$, $E_2$ $\to E$ one might wonder why the partner function indices are reversed for $E_1$ w.r.t. $E_2$ (or $E$ in \ctv) in \cref{eq:vb_spinors_dsh} (\cref{eq:vb_spinors} in the main text). The reason is simple and lies in our choice of matrix representations in \cref{app:irreps_ctv_dsh}, where the representative of $\sigma_{v1}$ has a different sign for $E_1$, reversing the $\sigma_{v1}$ parity characteristics.

Finally time reversal symmetry will bring further constraints on the UREFs, e.g. via $K\uline\Psi_1^{E_{1/2,u}}=\uline\Psi_2^{E_{1/2,u}}$ and $K\uline\Psi_2^{E_{1/2,u}}=-\uline\Psi_1^{E_{1/2,u}}$ due to Kramers degeneracy. Properties similar to \cref{eq:time_rev_req} can then be readily obtained, but are omitted for clarity.

{\begin{subequations}
\label{eq:vb_spinors_dsh}
\begin{equation}
\underline{\psi}_1^{E_{1/2,u}}=\frac{1}{\sqrt{2}}\left(\begin{array}{c}
\phi_1^{E_{1,u}}-\phi_2^{E_{2,u}}\\
\phi^{A_{1,u}}+\Phi_2^{E_{1,u}}\\
\phi^{A_{2,u}}+\Phi_1^{E_{1,u}}\\
\phi_2^{E_{1,u}}-\phi_1^{E_{2,u}}
\end{array}\right)
\label{eq:psi_1_E_12u}
\end{equation}
\begin{equation}
\underline{\psi}_2^{E_{1/2,u}}=\frac{1}{\sqrt{2}}\left(\begin{array}{c}
\phi_2^{E_{1,u}}+\phi_1^{E_{2,u}}\\
-\phi^{A_{2,u}}+\Phi_1^{E_{1,u}}\\
\phi^{A_{1,u}}-\Phi_2^{E_{1,u}}\\
-\phi_1^{E_{1,u}}-\phi_2^{E_{2,u}}
\end{array}\right)\label{eq:psi_2_E_12u}
\end{equation}
\begin{equation}
\underline{\psi}_1^{E_{3/2,u}}=\frac{1}{\sqrt{2}}\left(\begin{array}{c}
\phi^{A_{1,u}}-\phi^{B_{2,u}}\\
\Phi_1^{E_{1,u}}-\Phi_2^{E_{2,u}}\\
\Phi_2^{E_{1,u}}+\Phi_1^{E_{2,u}}\\
\phi^{A_{2,u}}+\phi^{B_{1,u}}
\end{array}\right)\label{eq:psi_1_E_32u}
\end{equation}
\begin{equation}
\underline{\psi}_2^{E_{3/2,u}}=\frac{1}{\sqrt{2}}\left(\begin{array}{c}
-\phi^{A_{2,u}}+\phi^{B_{1,u}}\\
\Phi_2^{E_{1,u}}-\Phi_1^{E_{2,u}}\\
-\Phi_1^{E_{1,u}}-\Phi_2^{E_{2,u}}\\
\phi^{A_{1,u}}+\phi^{B_{2,u}}
\end{array}\right)\label{eq:psi_2_E_32u}
\end{equation}
\begin{equation}
\underline{\psi}_1^{E_{5/2,u}}=\frac{1}{\sqrt{2}}\left(\begin{array}{c}
\phi_2^{E_{1,u}}-\phi_1^{E_{2,u}}\\
\phi^{B_{1,u}}+\Phi_2^{E_{2,u}}\\
\phi^{B_{2,u}}+\Phi_1^{E_{2,u}}\\
\phi_1^{E_{1,u}}-\phi_2^{E_{2,u}}
\end{array}\right)\label{eq:psi_1_E_52u}
\end{equation}
\begin{equation}
\underline{\psi}_1^{E_{5/2,u}}=\frac{1}{\sqrt{2}}\left(\begin{array}{c}
-\phi_1^{E_{1,u}}-\phi_2^{E_{2,u}}\\
-\phi^{B_{2,u}}+\Phi_1^{E_{2,u}}\\
\phi^{B_{1,u}}-\Phi_2^{E_{2,u}}\\
\phi_2^{E_{1,u}}+\phi_1^{E_{2,u}}
\end{array}\right)\label{eq:psi_2_E_52u}\end{equation}
\end{subequations}
}
%\input{supplemental.tex}
%\bibliography{refs_qd}% Produces the bibliography via BibTeX.%merlin.mbs apsrev4-1.bst 2010-07-25 4.21a (PWD, AO, DPC) hacked

\end{document}